\documentclass[twocolumn,preprintnumbers,amsmath,amssymb,aps,prl,showkeys,longbibliography]{revtex4-1}

\usepackage[T1]{fontenc}
\usepackage{graphicx}% Include figure files
\usepackage{dcolumn}% Align table columns on decimal point
\usepackage{bm}% bold math

\usepackage{yfonts}
\usepackage{float}
\usepackage{hyperref}
\usepackage{mathrsfs}
\usepackage{xcolor}
\usepackage{pgfplots}
\usepackage{bbold}
\usepackage{multibib}

\usepackage{filecontents}
\usepackage{amsmath}
\usepackage{mathtools}

\begin{document}

%\preprint{APS/123-QED}

\title{Deterministic single-atom source of quasi-superradiant $N$-photon pulses}% Force line breaks with \\

\author{Caspar Groiseau$^{1,2}$}
% \email{cgro288@aucklanduni.ac.nz}
\author{Alexander E. J. Elliott$^{1,2}$}
\author{Stuart J. Masson$^{3}$}
\author{Scott Parkins$^{1,2}$}
\email{s.parkins@auckland.ac.nz}
\affiliation{%
$^1$Dodd-Walls Centre for Photonic and Quantum Technologies, New Zealand\\
$^2$Department of Physics, University of Auckland, Auckland 1010, New Zealand\\
$^3$Department of Physics, Columbia University, 538 West 120th Street, New York, New York, 10027-5255, USA
}%

\date{\today}% It is always \today, today,

\begin{abstract}
We propose a single-atom, cavity quantum electrodynamics system, compatible with recently demonstrated, fiber-integrated micro- and nano-cavity setups, for the on-demand production of optical number-state, $0N$-state, and binomial-code-state pulses. The scheme makes use of Raman transitions within an entire atomic ground-state hyperfine level and operates with laser and cavity fields detuned from the atomic transition by much more than the excited-state hyperfine splitting. This enables reduction of the dynamics to that of a simple, cavity-damped Tavis-Cummings model with the collective spin determined by the total angular momentum of the ground hyperfine level.
\end{abstract}

   
%\keywords{Quantum Optics, Quantum Information, Cold Atoms}%Use showkeys class option if keyword
                             
\maketitle

\textit{Introduction.}--Recent experiments with trapped atoms and fiber-integrated, optical micro- and nano-cavities have pushed the field of cavity quantum electrodynamics (cavity QED) into a new realm of single-atom--photon coupling strengths, corresponding to unprecedentedly large single-atom cooperativities \cite{PhysRevLett.104.203602,Haas180,Barontini:2015aa,thompson2013coupling,tiecke2014nanophotonic,PhysRevLett.124.063602}, while also offering the possibility of integrated quantum networks for quantum communication or simulation of quantum many-body systems \cite{tan2019resurgence,slussarenko2019photonic,flamini2018photonic,o2009photonic,kok2007linear,kimble2008quantum,o2007optical}.
A further, well-known capability provided by such large coupling strength is the generation of single photons with high fidelity through cavity-enhanced atomic spontaneous emission. Efficient single-photon sources are of course central to many efforts to realize optical quantum computation and communication.

Beyond this, however, lies an even greater, and still outstanding challenge to produce a similarly efficient source of pulses containing {\em exactly} $N$ ($\geq 2$) optical photons. Such highly nonclassical states of light are of fundamental interest to quantum optics and constitute a starting point for the engineering of yet more complex quantum states. They are also essential for newly-emerging, more resource-efficient photonic architectures for universal quantum computation and quantum error correction using individual, higher-dimensional systems  \cite{PhysRevA.97.062315,PhysRevX.6.031006,hu2019quantum} ({\em cf}. multiple two-state systems), as well as for optimal capacity of a quantum communication channel \cite{RevModPhys.58.1001,RevModPhys.66.481}, and Heisenberg-limited quantum metrology (interferometry) \cite{slussarenko2017unconditional,RevModPhys.84.777,nagata2007beating,mitchell2004super,PhysRevLett.110.163604,PhysRevA.68.023810,PhysRevLett.71.1355}.

Recent proposals and proof-of-principle demonstrations of optical $N$-photon sources, typically using parametric down conversion or quantum dots, are intrinsically probabilistic and low-yield in nature \cite{munoz2014emitters,1367-2630-8-1-004,krapick2016chip,Moebius:16,fischer2017signatures,khoshnegar2017solid,loredo2019generation}. The use of a known number of (effective) two-level atoms emitting into a cavity or photonic waveguide has also been proposed  \cite{PhysRevA.67.043818,PhysRevLett.115.163603,PhysRevLett.118.213601,PhysRevA.99.043807}, but the required many-body control and repeatability is still very challenging. 

Here, we propose a deterministic $N$-photon source that requires just a {\em single} atom and makes use of its entire, multilevel energy structure in a manner that reduces the effective system dynamics to a simple and transparent form. In particular, we demonstrate that a single alkali atom coupled strongly to a cavity mode and subject to Raman transitions between sublevels of a ground $F$ hyperfine state can replicate the collective emission of $N=2F$ initially excited, two-level atoms into the cavity mode. In this way, a ``superradiant'' pulse of precisely $N$ photons can be extracted, through the cavity mode, from a {\em single} atom. Moreover, an initial, coherent superposition state of the atom's ground sublevels is also preserved in the emission process, enabling the generation of a light pulse in an arbitrary superposition of Fock states; as particular examples, we consider $0N$-state and binomial-code-state pulses of light.

Key to the reduction of the single-atom, multilevel dynamics is a very large detuning of the laser and cavity fields from an entire excited state hyperfine manifold of the atom, such that the excited-state hyperfine splittings can be ignored; alternatively, such that the total electronic angular momentum, $J$, is a good quantum number. Such large detuning from the atomic transition in turn demands a very large atom-cavity coupling strength and, as we show here, the experimental configurations of \cite{PhysRevLett.104.203602,Haas180,Barontini:2015aa,thompson2013coupling,tiecke2014nanophotonic,PhysRevLett.124.063602} attain the requisite strength for our scheme to be feasible and efficient.

The potential for making use of the multilevel energy structure of an alkali atom to prepare $N$-photon states has been considered previously, using either adiabatic passage with time-dependent laser and atom-cavity coupling strengths \cite{ParkinsPRL1993,ParkinsPRA1995}, or cavity-mediated optical pumping between atomic ground state sublevels \cite{PhysRevA.86.063801}. However, in contrast to the present scheme, these approaches assume near-resonant laser and cavity fields and consider just a single $F\leftrightarrow F'$ transition. This limits the range of validity of the approaches and means that Clebsch-Gordan coefficients between $m_F$ and $m_{F'}$ sublevels play a nontrivial and restricting (with regards to choice of $F$ and $F'$) role in the scope and performance of the scheme. 

\begin{figure}[H]
    \begin{center}
	\includegraphics[width=0.9\linewidth]{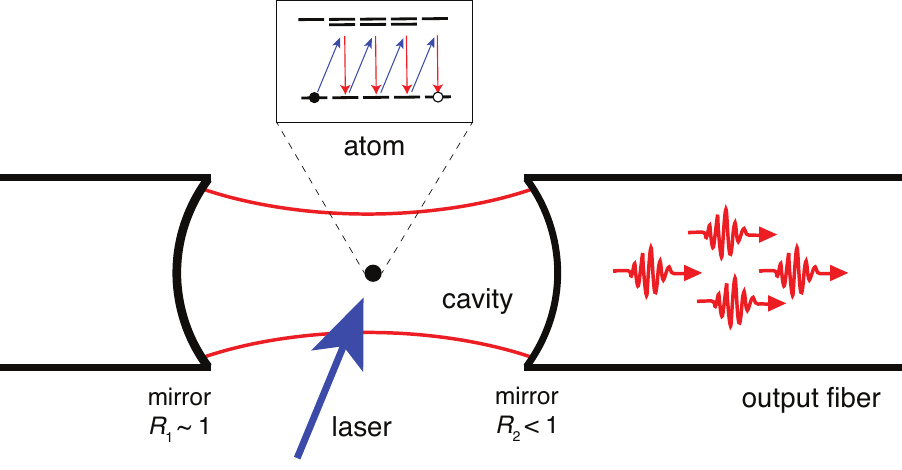}
	\end{center}
	\caption{Fiber-cavity configuration with a $\sigma_+$-polarized laser and $\pi$-polarized cavity mode coupled to the $D_1$ line of a ${}^{87}{\rm Rb}$ atom. The atom is initially prepared in the $\{ F=2,m_F=-2\}$ ground-state sublevel. The cavity field decays predominantly through the right-hand mirror.}
\label{schemeplot}	
\end{figure}

\textit{Engineered Tavis-Cummings Dynamics.}--We consider a single alkali atom tightly confined inside an optical cavity. The atom couples to a $\pi$-polarized cavity mode (annihilation operator $\hat a$) and is also driven by either a $\sigma_+$- or $\sigma_-$-polarized laser field (Fig.~\ref{schemeplot}). We define the atomic dipole transition operators
\begin{align}
&\hat D_q(F,F') = \nonumber \\
& \sum_{m_F=-F}^{F} |F,m_F\rangle\langle F,m_F|\mu_q|F',m_F+q\rangle\langle F',m_F+q| ,    
\end{align}
where $q=\{ -1,0,1\}$ and $\mu_q$ is the dipole operator for $\{\sigma_-,\pi,\sigma_+\}$-polarization, normalized such that $\langle\mu\rangle =1$ for a cycling transition.
The master equation for the density operator, $\hat\rho$, of our system in the interaction picture is ($\hbar=1$)
\begin{equation}
\label{fullME}
    \dot{\hat\rho}=-i\left[\hat H_\pm,\hat\rho\right]+ \kappa\mathcal{D}[\hat a]\hat\rho
    +\frac{\gamma}{2} \sum_q \mathcal{D} \left[ \sum_{F,F'} \hat D_q(F,F') \right] \hat\rho ,
\end{equation}
where $\kappa$ is the cavity field decay rate, $\gamma$ the free-space atomic spontaneous emission rate, and
$\mathcal{D}[\hat O]\hat\rho=2\hat O\hat\rho\hat O^\dagger-\hat\rho\hat O^\dagger\hat O-\hat O^\dagger\hat O\hat\rho$. 
Setting the zero of energy at the lower ground hyperfine level, and assuming (for the moment) zero magnetic field,
the Hamiltonian is
\begin{equation}
\label{fullH}
\begin{split}
    &\hat H_\pm=\Delta_\text{c}\hat a^\dagger \hat a+\sum_{m_{F_\uparrow}} \omega_\text{GHS} |F_\uparrow ,m_{F_\uparrow}\rangle\langle F_\uparrow ,m_{F_\uparrow}|\\
    &-\sum_{F',m_{F'}} \Delta_{F'} |F',m_{F'}\rangle\langle F',m_{F'}|\\
    &+\sum_{F,F'}\left( \frac{\Omega}{2} \hat D_{\pm 1}(F,F')  
    +g\hat a^\dagger \hat D_0(F,F') +\text{H.c.} \right) .
\end{split}
\end{equation}
Here, $\Delta_\text{c}=\omega_\text{c}-\omega_\pm$ is the detuning between the cavity and laser frequencies, $\Omega=|\Omega|e^{i\phi}$ the Rabi frequency of the $\sigma_\pm$-polarized laser field, $g$ the atom-cavity coupling strength, $\omega_\text{GHS}$ the ground state hyperfine splitting [$F_\uparrow$ ($F_\downarrow$) denotes the upper (lower) hyperfine ground state], and $\Delta_{F'}=\omega_\pm-\omega_{F'}$ the detuning of the laser from the $F_\downarrow \leftrightarrow F'$ transition.
Note that, given the large coupling strengths and detunings that we consider here, we assume that the light fields couple all hyperfine ground and excited states. Consistent with this, we also assume that all atomic decays of a given polarization are into a common reservoir \cite{BirnbaumPRA2006}.

If we now assume also, more specifically, that the detunings of the fields (cavity and laser) are much larger than the excited state hyperfine splitting, such that this splitting can essentially be neglected, then, in addition to being able to adiabatically eliminate the atomic excited states and neglect atomic spontaneous emission, we obtain a tremendously simplified effective model of the atom-cavity dynamics in the form of an anti-Tavis-Cummings or Tavis-Cummings model (anti-TCM or TCM, depending on the polarization of the laser field) for a collective spin $F$ \cite{PhysRevLett.119.213601,zhiqiang2017nonequilibrium,PhysRevLett.122.103601}, where $F$ (either $F_\uparrow$ or $F_\downarrow$) is determined by the initial state of the atom; i.e., our master equation reduces to
\begin{equation}
\label{effME}
    \dot{\hat\rho}=-i\left[\hat{\mathcal{H}}_\pm,\hat\rho\right]+ \kappa\mathcal{D}[\hat a]\hat\rho ,
\end{equation}
with
\begin{equation}
\label{effH}
 \hat{\mathcal{H}}_\pm=\omega\hat a^\dagger \hat a+\omega_0\hat S_z+\lambda\left(e^{-i\phi}\hat a\hat S_{\mp}+ e^{i\phi}\hat a^\dagger\hat S_{\pm}\right) ,
\end{equation}
where $\{ \hat S_{\pm},\hat S_z\}$ are the spin-$F$ angular momentum operators and, for example, for the $D_1$ line of ${}^{87}{\rm Rb}$, the effective parameters are
\begin{equation}
	\omega=\Delta_\text{c}+\frac{g^2}{3\Delta} ,~~~
	\omega_0=\omega_\text{Z}\mp\frac{|\Omega|^2}{24\Delta} , ~~~
	\lambda=\frac{g|\Omega|}{12\sqrt{2}\Delta}.
\end{equation}
Here, we now assume an external magnetic field giving rise to a shift $\omega_\text{Z}$ of the $m_F$ levels. The detuning $\Delta$ depends on the choice of $F$; for $F=F_\uparrow$, we take $\Delta=\Delta_{F'}+\omega_\text{GHS}$, where the choice of $F'$ in $\Delta_{F'}$ makes little difference due to the very large detuning assumed (in practice, we pick the lowest $F'$). The same forms of expressions for $\{\omega,\omega_0,\lambda\}$ are obtained for the $D_1$ and $D_2$ lines of other alkali atoms, but with slightly different numerical factors.

The essence of our scheme follows clearly and simply from the dynamics described by (\ref{effME}) and (\ref{effH}). With the choice $\hat{\mathcal{H}}_\pm$ and corresponding initial atom-cavity state $|F,m_F=\mp F\rangle |0\rangle_{\rm cav}$, the system evolves irreversibly to the state $|F,m_F=\pm F\rangle |0\rangle_{\rm cav}$ with emission from the cavity of a pulse of exactly $2F$ photons. Additionally, in the regime of interest to us, where $\kappa\gg\sqrt{F}\lambda$ and $\omega\simeq\omega_0\simeq 0$ (via tuning of $\Delta_\text{c}$ and $\omega_\text{Z}$), the effective model describes resonant, cavity-mediated superradiant emission of the collective spin; i.e., the emitted $2F$-photon pulse will have a characteristic $\text{sech}^2$-shaped temporal profile \cite{doi:10.1119/1.1986858}.

\textit{Output photon number.}--A preliminary way to quantify the quality of our state generation scheme is to compute the time evolution of the output photon flux from the cavity and the mean number of emitted photons, $\overline{N}=2\kappa\int_0^\infty dt\langle \hat a^\dagger (t) \hat a (t)\rangle$. 
We consider first the case of a ${}^{87}{\rm Rb}$ atom initially prepared in the ground state $|F=2,m_F=-2\rangle$ and coupled to the laser and cavity fields via the $D_1$ line. With this system, we expect an output pulse of exactly 4 photons.

\begin{center}

\begin{figure}[t]
\centering
	\includegraphics[width=0.49\linewidth]{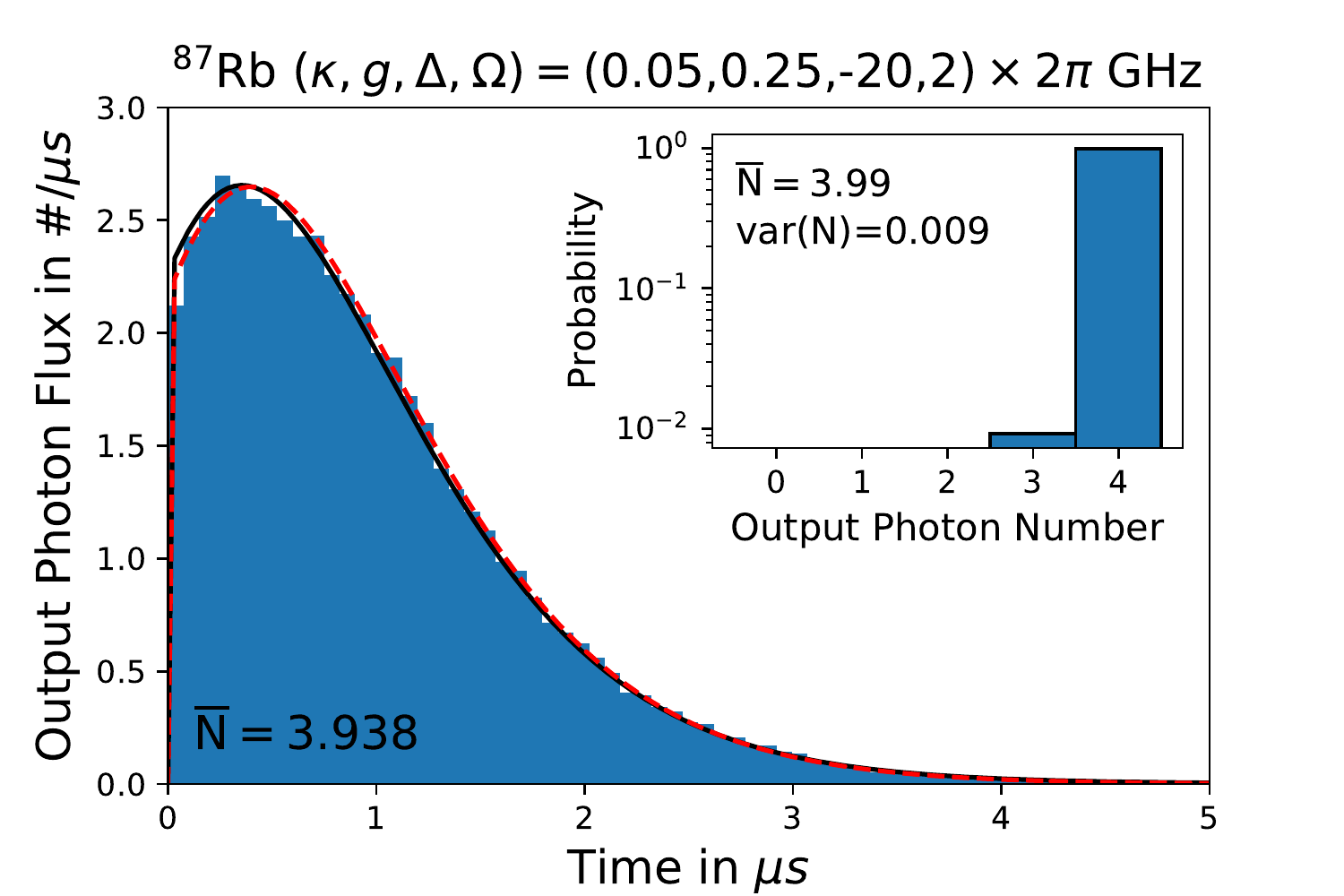}
	\includegraphics[width=0.49\linewidth]{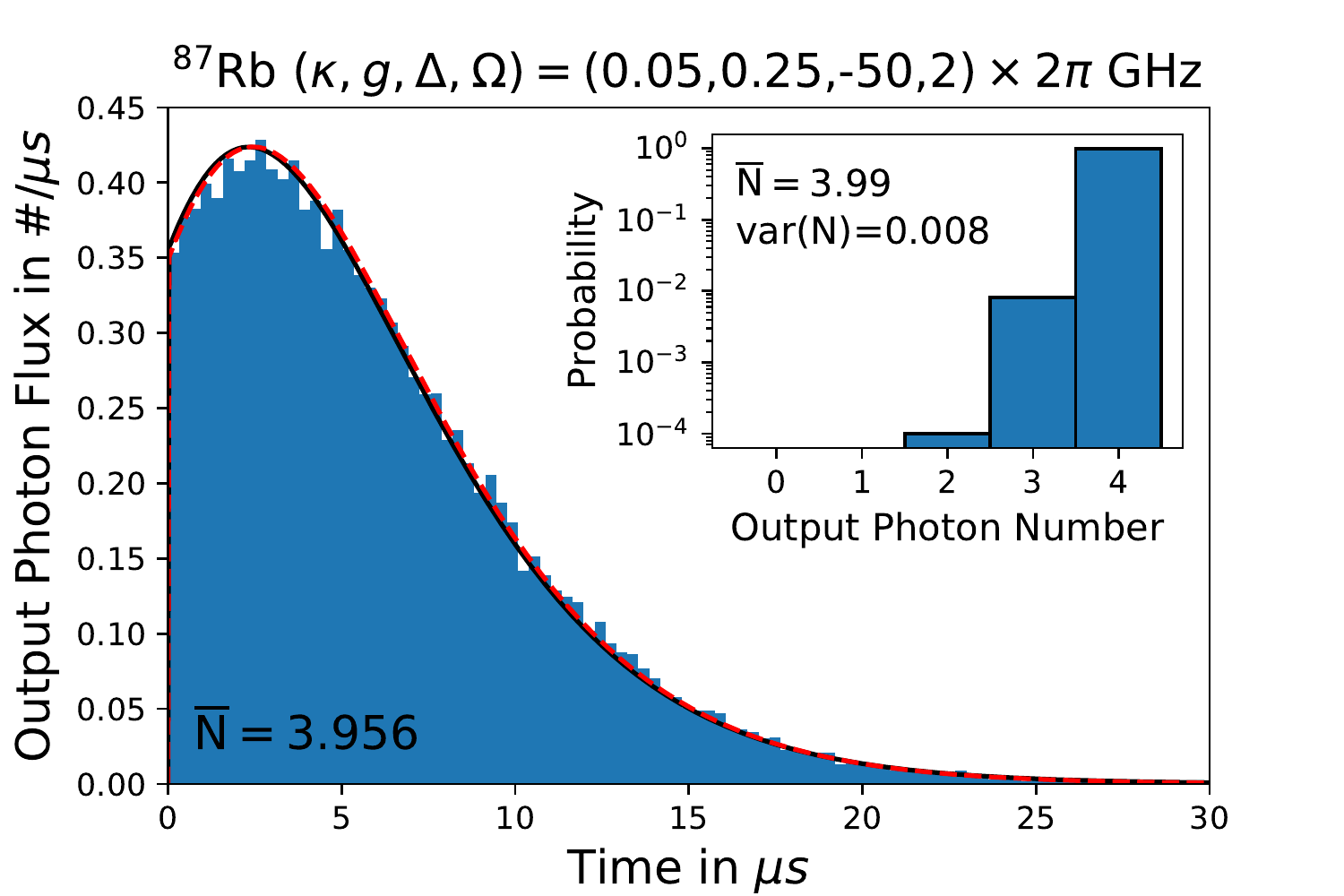}\\
	\includegraphics[width=0.49\linewidth]{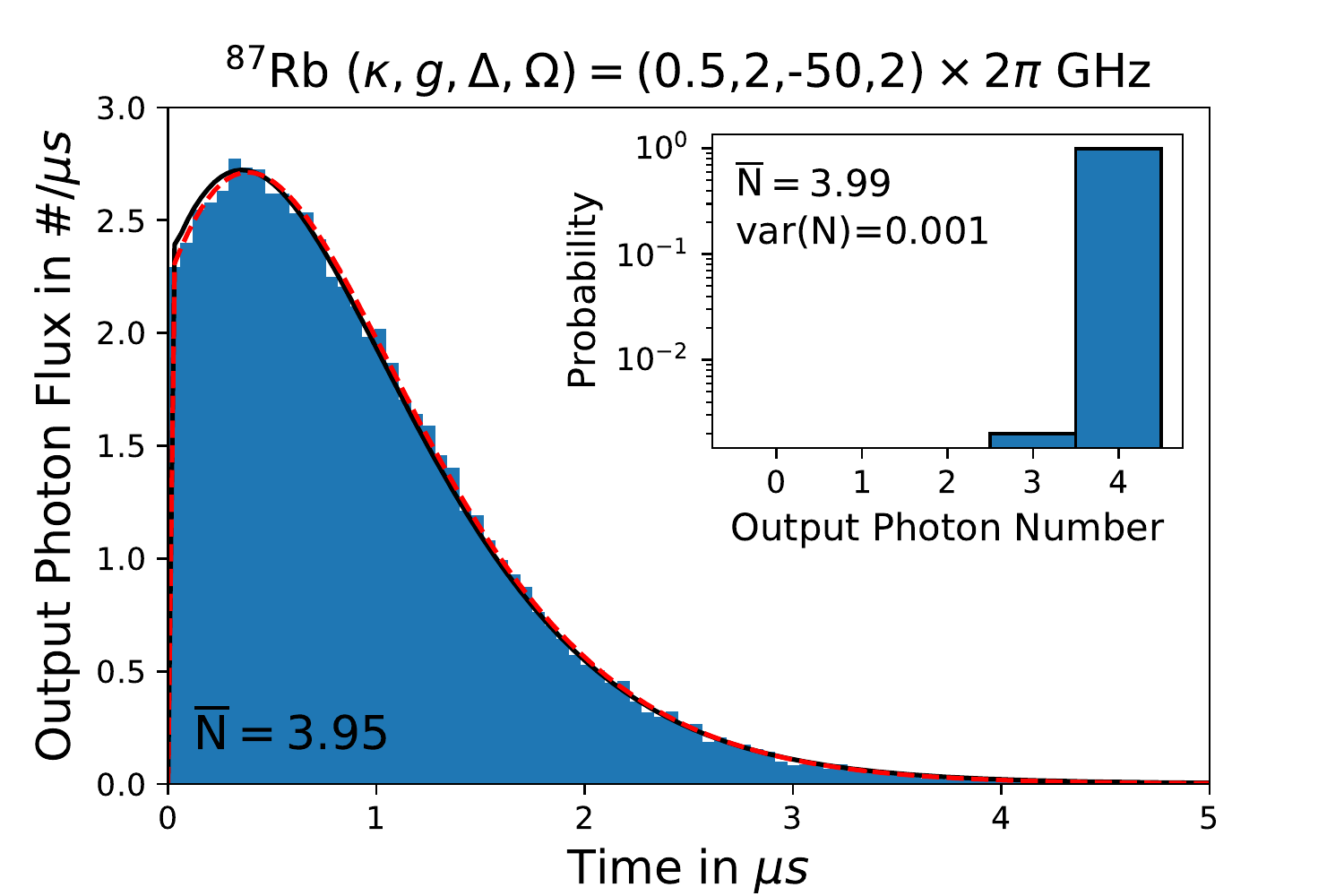}
	\includegraphics[width=0.49\linewidth]{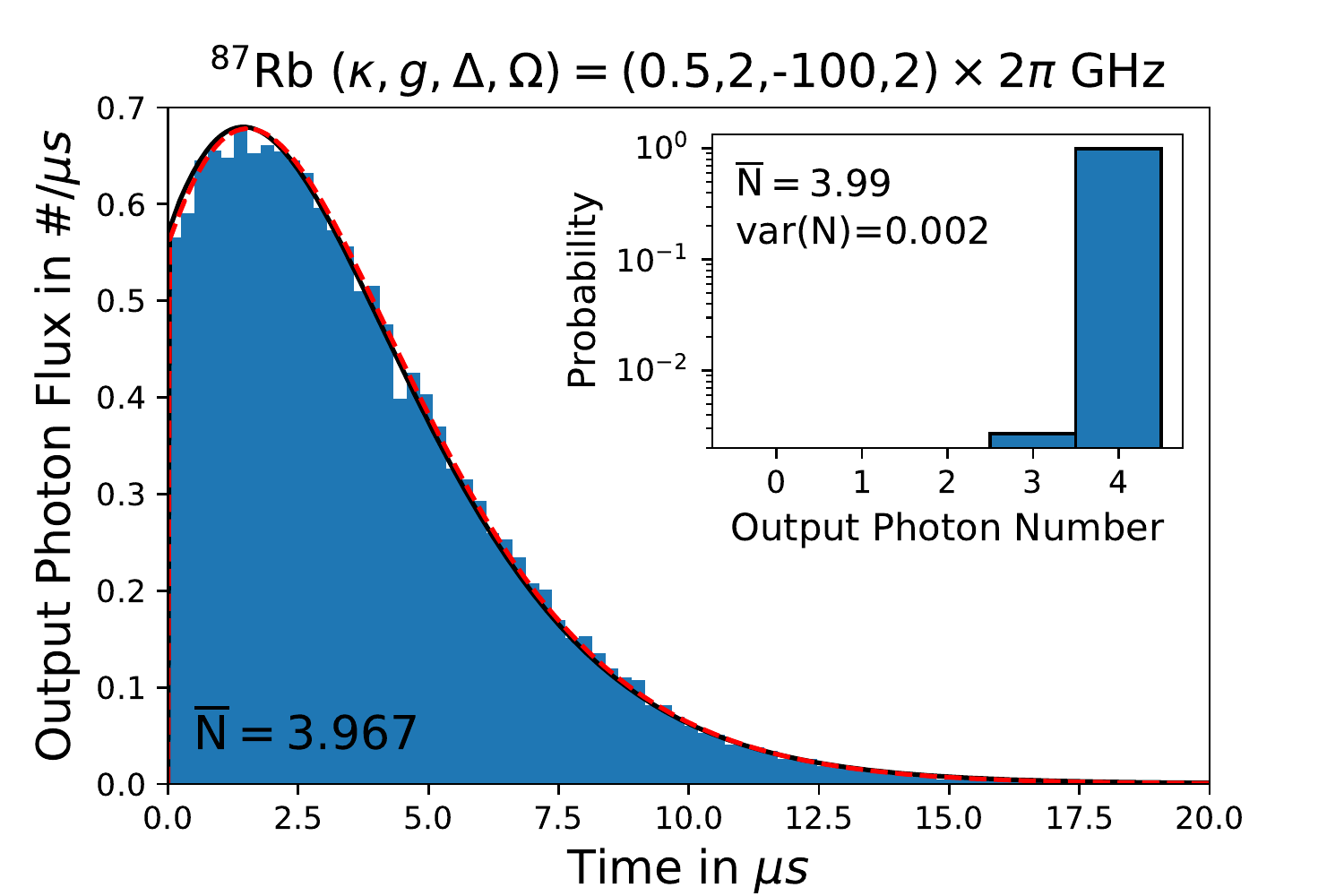}
	\caption{Output photon flux for a ${}^{87}{\rm Rb}$ atom initially prepared in $|F=2,m_F=-2\rangle$. The black solid line represents the full model and the red dashed line the anti-TCM. The histogram shows the temporal distribution of photocounts (renormalised to $\bar{N}$) for 10000 trajectories of the anti-TCM with additional, effective spontaneous emission. The number below the curves gives $\bar{N}$ for the full model. Insets: Histogram of photon number counts per trajectory (output pulse).}
	\label{Rbplots}
\end{figure}
\end{center}

We solve the master equation numerically for the full model, (\ref{fullME}--\ref{fullH}), and compare results with the solution for the effective anti-TCM, (\ref{effME}--\ref{effH}), for two sets of cavity QED parameters: (i) $\{\kappa,g,\gamma\}/2\pi=\{ 50,250,5.7\}\text{ MHz}$ and (ii) $\{\kappa,g,\gamma\}/2\pi=\{ 0.5,2,0.0057\}\text{ GHz}$. The first set corresponds to the fiber microcavity system of \cite{PhysRevLett.104.203602,Haas180,Barontini:2015aa}, while the second set is relevant to the nanocavity system of \cite{thompson2013coupling,tiecke2014nanophotonic,PhysRevLett.124.063602}.
The results are shown in Fig.~\ref{Rbplots}, with two different detunings $\Delta$ considered for each set. The agreement between the full and reduced models is clearly very good, and the predicted $\text{sech}^2$-shaped pulse is confirmed, with a duration on the order of $(F\lambda^2/\kappa )^{-1}$. The atomic state populations in the full model also show the expected evolution, with a smooth transfer of population along the $F=2$ hyperfine level to the final state $|F=2,m_F=+2\rangle$ (see SI). A very small fraction of population may be transferred to the $F=1$ ground state via off-resonant processes, but, in fact, the effective superradiant emission simply continues from within this level and any population there is ultimately driven back (also by an off-resonant process) into the $F=2$ level and so to the final (dark) state $|F=2,m_F=+2\rangle$.

\begin{center}
\begin{figure}[b]
\centering
	\includegraphics[width=0.49\linewidth]{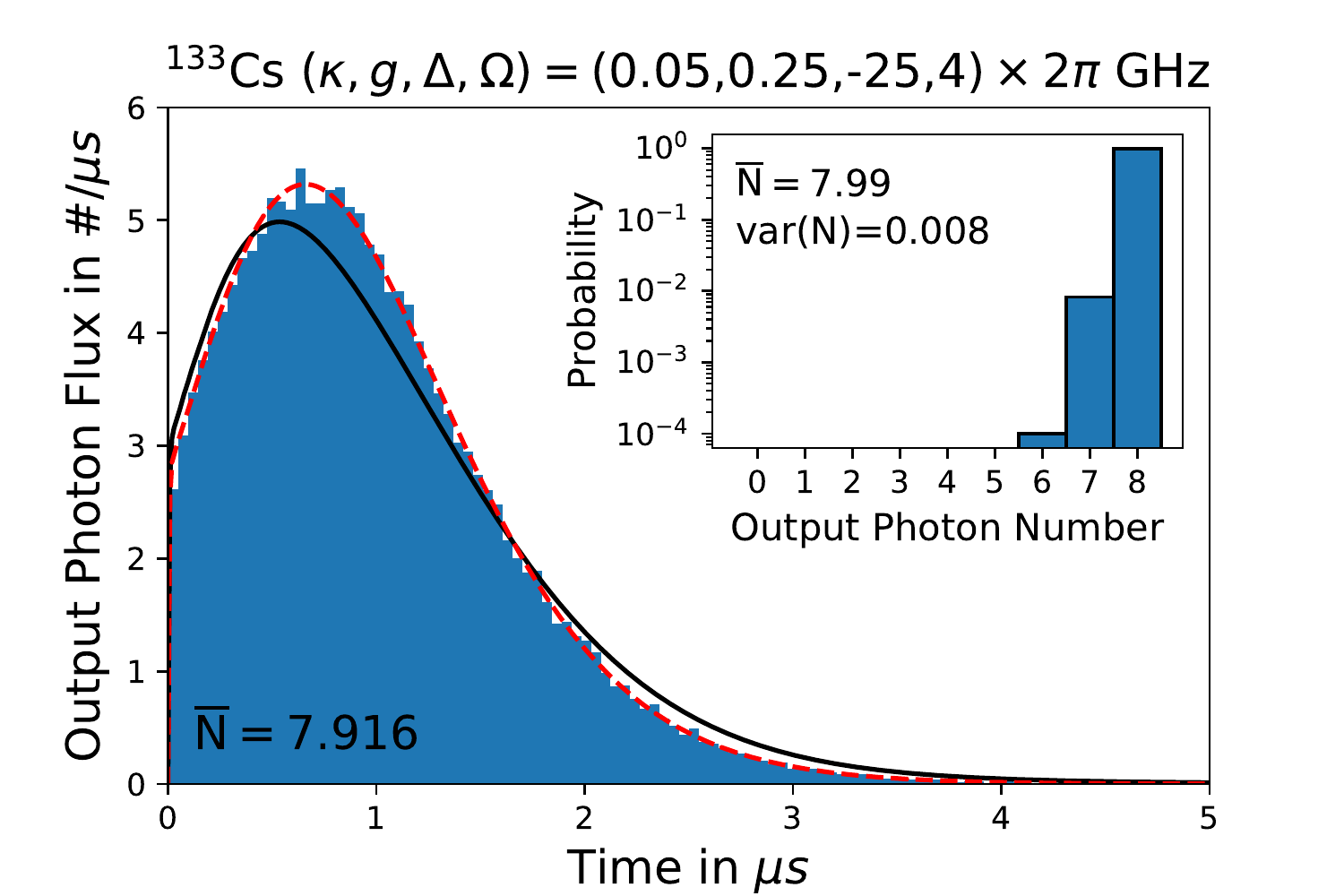}
	\includegraphics[width=0.49\linewidth]{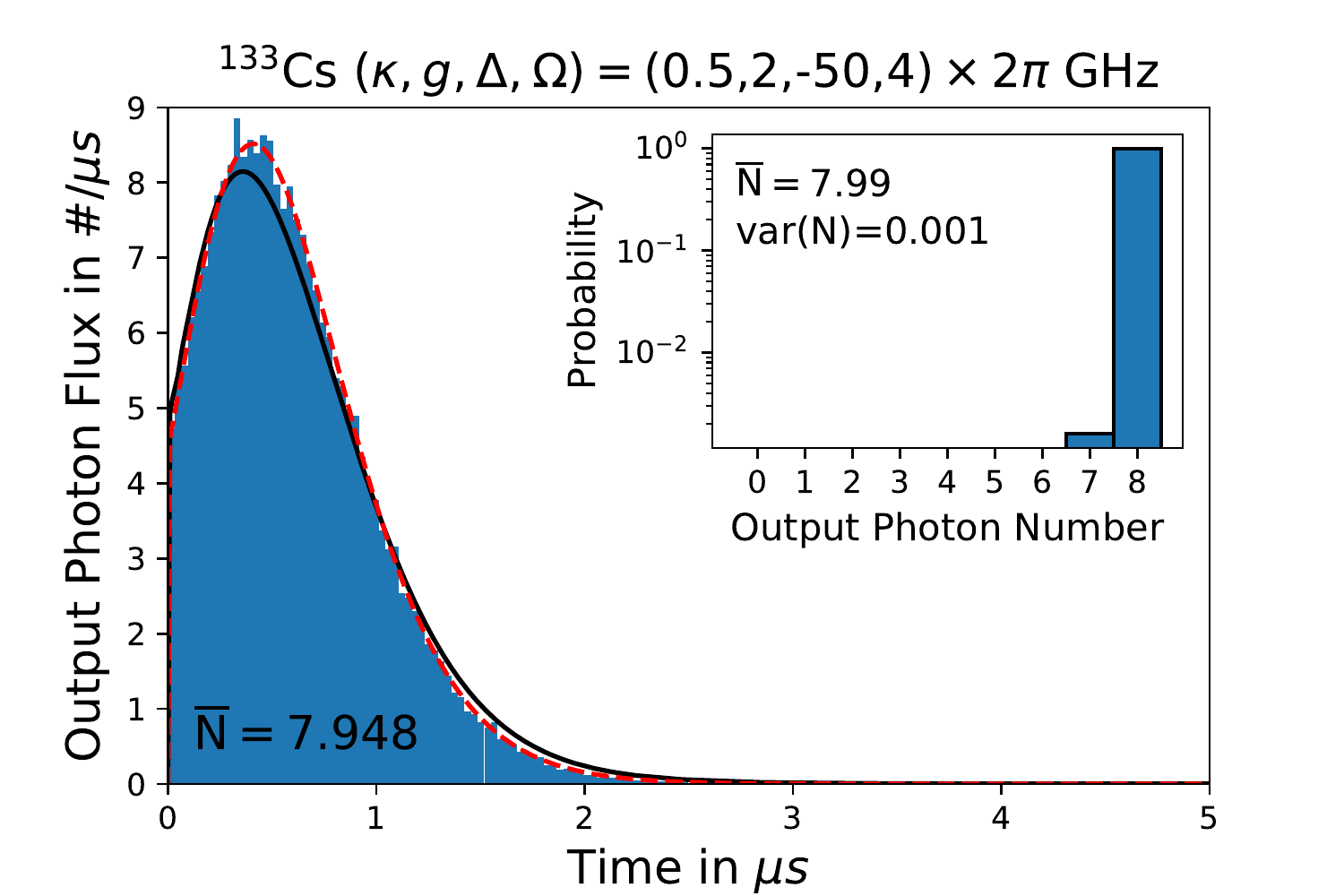}
	\caption{Output photon flux for a ${}^{133}{\rm Cs}$ atom initially prepared in $|F=4,m_F=-4\rangle$. The black solid line represents the full model and the red dashed line the anti-TCM. The histogram shows the temporal distribution of photocounts (renormalised to $\bar{N}$) for 10000 trajectories of the anti-TCM with additional, effective spontaneous emission. The number below the curves gives $\bar{N}$ for the full model. Insets: Histogram of photon number counts per trajectory (output pulse).}
	\label{Csplots}
\end{figure}
\end{center}

Similar results are shown in Fig.~\ref{Csplots} for a ${}^{133}{\rm Cs}$ atom initially prepared in the $|F=4,m_F=-4\rangle$ ground state, also operating on the $D_1$ line. For the parameters used, there is a slight discrepancy between pulse shapes for the two models, but the mean photon number obtained from the full model is still very close to the expected value of 8. 
The discrepancy arises primarily from the larger excited state hyperfine splitting in ${}^{133}{\rm Cs}$, which means that a larger detuning is required to ensure closer agreement with the TCM; a similar discrepancy is also observed in ${}^{87}{\rm Rb}$ for smaller detunings (see SI).

The integrated photon flux obtained from the master equation, however, does not tell us about the variance in the photon number of the output pulse. To get the variance we perform quantum trajectory simulations \cite{carmichael1993open,Molmer:93} and record the times and total number of photon counts in each trajectory. The histogram of photon detection times gives us again the output photon flux, which is shown for comparison in Figs.~\ref{Rbplots} and \ref{Csplots}, along with the photon number distribution and its variance for the output pulse. The distributions are clearly very close to an ideal number state. Note that for these simulations we do not use the full model (owing to the stiffness of the numerical integration caused by the large detunings and ground state hyperfine splitting), but rather use the anti-TCM (or TCM) with spontaneous emission added, in the form of an effective Lindblad operator acting just within the relevant ground state (see SI). 
Spontaneous emission to the other hyperfine ground state is therefore neglected, but the numerical results from the master equation support this as a good approximation.

Finally, we note that shorter or longer output pulses can be obtained by changing the detuning $\Delta$ and/or laser Rabi frequency $\Omega$. Also, in the results presented here, we assume an instantaneous switch-on of the laser field. This can be relaxed to allow for a smooth initial ramp of the laser field to its peak value, which can be used to tailor the shape of the output $2F$-photon pulse (see the SI for additional examples).

\twocolumngrid

%\onecolumngrid
\twocolumngrid

\begin{center}
\begin{figure}[t]
\centering
	\includegraphics[width=0.49\linewidth]{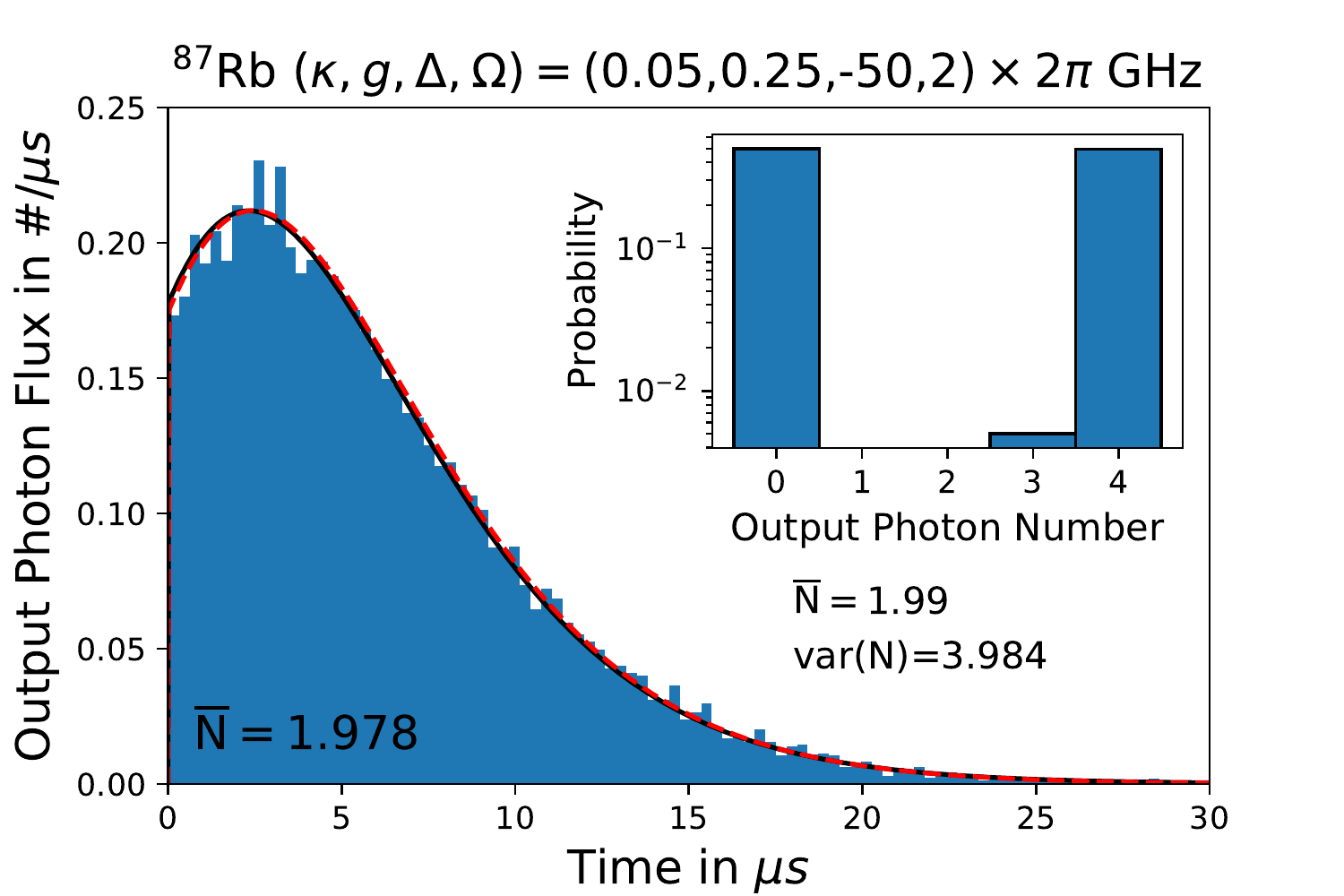}
	\includegraphics[width=0.49\linewidth]{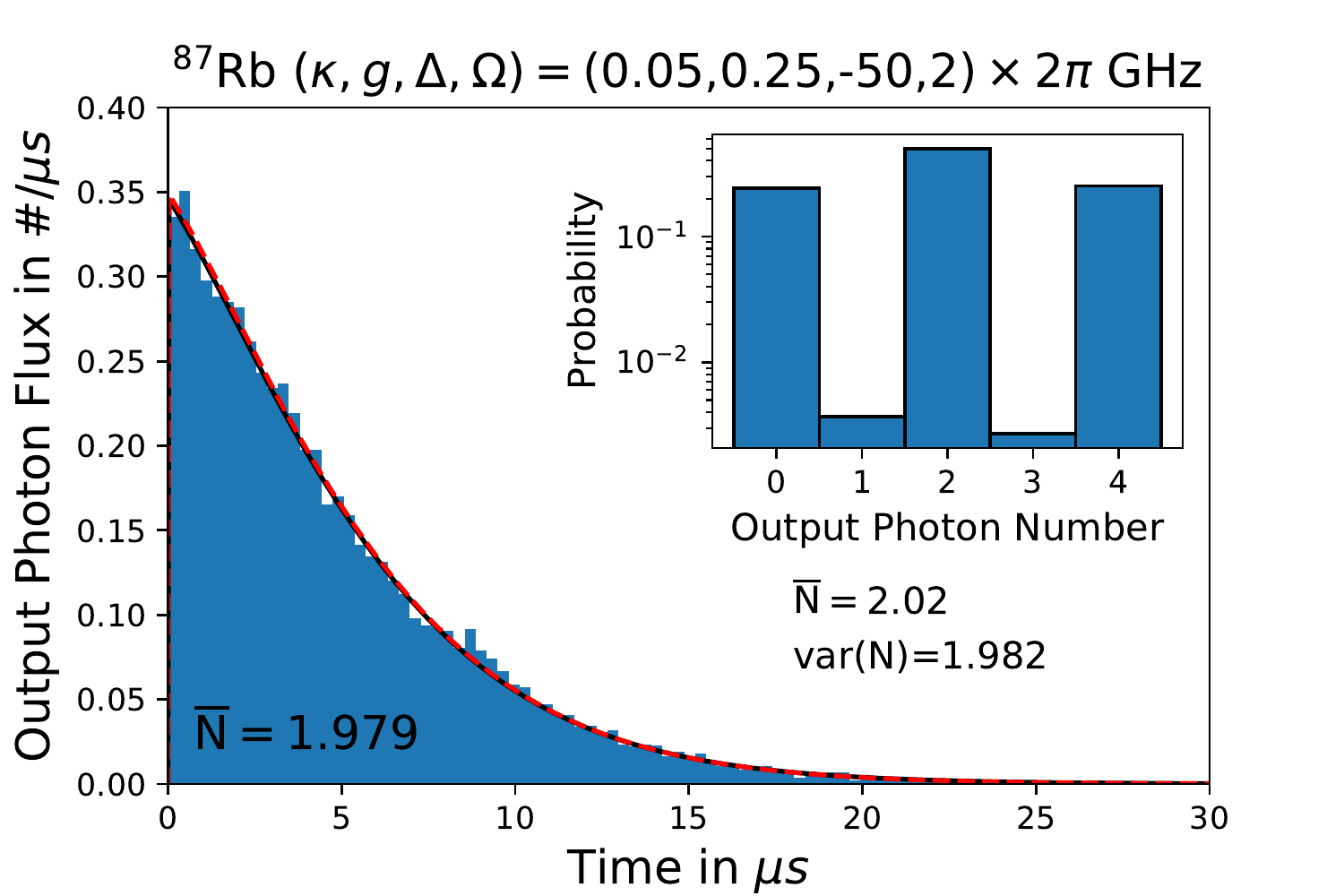}
	\caption{Output photon flux for a ${}^{87}{\rm Rb}$ atom initially prepared in the states $|\psi_\text{oat}\rangle_{F=2}$ (left) and $|\psi_\text{bc}\rangle_{F=2}$ (right). The black solid line represents the full model and the red dashed line the anti-TCM. The histogram shows the temporal distribution of photocounts (renormalised to $\bar{N}$) for 10000 trajectories of the anti-TCM with additional, effective spontaneous emission. The number below the curves gives $\bar{N}$ for the full model. Insets: Histogram of photon number counts per trajectory (output pulse).}
	\label{superpositionplots}
\end{figure}
\end{center}

\textit{$0N$-states and other superpositions.}--Instead of starting with an atom in an end state of the Zeeman ladder, we also have the option to start with a superposition of Zeeman substates; for example, $|\psi_\text{oat}\rangle_F=(|F,-F\rangle+|F,+F\rangle )/\sqrt{2}$, which can be created using a one-axis twisting (oat) scheme \cite{chalopin2018quantum}, or an arbitrary superposition, created using a scheme such as in \cite{law1998synthesis}. With our proposed system, these atomic superposition states are directly mapped onto photonic states of the output light pulse. So, for example, the initial state $|\psi_\text{oat}\rangle_F$ leads to an output pulse in a coherent superposition of vacuum and $N=2F$ photons, i.e., an $0N$-state, $|\Psi\rangle_{\text{pulse}}=(|0\rangle_\text{out}+|N\rangle_\text{out})/\sqrt{2}$, which is a basic resource in schemes proposed for universal quantum computation \cite{PhysRevA.97.062315}. The output photon flux and photon number distribution for initial atomic state $|\psi_\text{oat}\rangle_{F=2}$ are shown in Fig.~\ref{superpositionplots} for ${}^{87}{\rm Rb}$ with cavity QED parameters relevant to the fiber microcavity.

As a further example, in Fig.~\ref{superpositionplots} we also consider an initial state of ${}^{87}{\rm Rb}$ of the form $|\psi_\text{bc}\rangle_{F=2}=(|2,-2\rangle+\sqrt{2}|2,0\rangle+|2,+2\rangle )/2$, yielding an output pulse state $|\Psi\rangle_{\text{pulse}}=(|0\rangle_\text{out}+\sqrt{2}|2\rangle_\text{out}+|4\rangle_\text{out})/2$. Such a state is of particular interest, as it constitutes a superposition of states $|0_\text{L}\rangle=(|0\rangle+|4\rangle )/\sqrt{2}$ and $|1_\text{L}\rangle=|2\rangle$, which are logically encoded (binomial code) states of a qubit for a quantum computation scheme protected up to one photon loss \cite{PhysRevX.6.031006}.
Note that in mapping general atomic ground-state superpositions onto the states of the output light pulses, one must pay attention to the relative phases between the different components and the phase $\phi$ of the effective cavity-spin coupling. For an exact mapping of relative phases, we require in our model that $\phi=\mp\pi/2$ (depending on the sign of $\Delta$), which can of course be chosen through the phase of the laser field (see SI for further details). Alternatively, $\phi$ can also be incorporated as the relative phase between neighbouring $m_F$ levels in the initial atomic state.

\twocolumngrid

\begin{center}
\begin{figure}[t]
\centering
\label{Wignerplots}
\includegraphics[width=\linewidth]{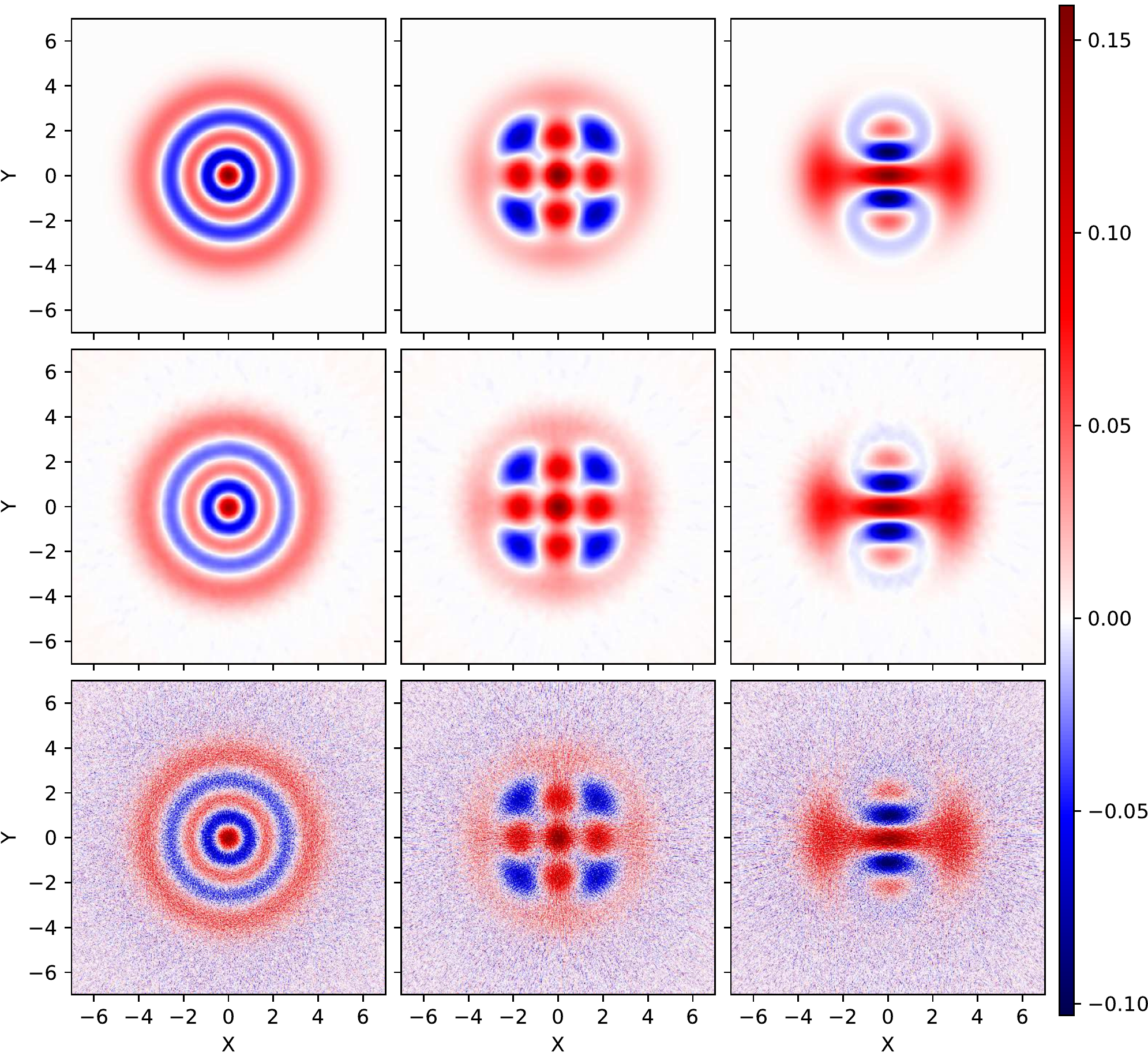}
	\caption{Top row: Wigner functions of the states (left to right) $|4\rangle$, $(|0\rangle+|4\rangle)/\sqrt{2}$, and $(|0\rangle+\sqrt{2}|2\rangle+|4\rangle)/2$. Bottom row: raw reconstructed Wigner functions of the cavity output pulses for a single ${}^{87}{\rm Rb}$ atom in a fiber microcavity setup ($\{\kappa,g,\Delta,\Omega\}=\{0.05,0.25,-50,2\}\cdot2\pi\text{ GHz}$) with initial atomic states (left to right) $|2,-2\rangle$, $|\psi_\text{oat}\rangle_{F=2}$, and $|\psi_\text{bc}\rangle_{F=2}$. The reconstructions are using a set of 500 angles $\theta\in[0,\pi)$ and $10000$ trajectories per angle. Middle row: Simulated reconstructions smoothed by a Gaussian blur.}
	\label{wigner}
\end{figure}	
\end{center}

\textit{Quantum state tomography.}--The photon number distribution of the output pulse is not sufficient to confirm that the desired output state has been generated. To verify that the target quantum state has indeed been generated, we implement quantum state tomography on simulated, pulsed-homodyne measurements, obtained via the method of homodyne quantum trajectories (see SI for details). That is, we reconstruct the Wigner function of the pulse by measuring marginals of the Wigner function for a set of homodyne phase angles $\theta$ and then applying the inverse Radon transform to these marginals \cite{PhysRevA.40.2847}. For these simulations, we again use the TCM with effective spontaneous emission added. Results of these reconstructions are shown in Fig.~\ref{wigner}. Our reconstructions can clearly be assigned to the predicted, ideal state. For a better comparison, we remove most of the noise from the simulated results by smoothing with a Gaussian blur, which reveals some discrepancy in the heights of the maxima and minima between the ideal case and our reconstruction. This effect can also be observed in the untreated marginals, where we observe some noise in the outer peaks of these marginals that can be attributed to atomic spontaneous emission.

Alternatively, the density matrix itself can be reconstructed using maximum likelihood estimation \cite{PhysRevA.55.R1561,Lvovsky2004} on the marginals. We start with a normalized identity matrix as the initial guess and go through $500$ iterations. The resulting fidelities of the estimated density matrices for the Fock state, $0N$-state and the binomial-code state from Fig. \ref{wigner} are $0.922$, $0.959$, and $0.968$, respectively. Almost identical fidelities have also been obtained using the input-output formalism for quantum pulses \cite{PhysRevLett.123.123604} (see SI). This illustrates that, even though the superposition states are somewhat more complex, the initial atomic superposition states are, in a certain sense, closer to the final atomic state, and therefore more robust against spontaneous emission.

 \textit{Conclusion and outlook.}--We have proposed a single-atom, deterministic source of optical number-state, $0N$-state, and binomial-code-state pulses. The scheme does not require time-dependent atom-laser or atom-cavity coupling strengths or detunings, or specific $F\leftrightarrow F'$ atomic transitions, 
 and should be feasible with recently-demonstrated, fiber-integrated micro- and nano-cavity QED setups. Some other potential features of the scheme are worth noting. For the case of number-state pulses, it is a simple matter to generate a stream of such pulses by simply switching the polarization of the laser field at the end of each pulse and cycling the atom back and forth between the end states $|F,\pm m_F\rangle$. In this context, one may also increase $N$ by adding more atoms; e.g., with two identically-prepared ${}^{87}{\rm Rb}$ atoms coupled collectively to the cavity mode, the effective spin in the TCM is simply doubled, enabling the generation of 8-photon pulses.
Finally, we have assumed throughout this work that the cavity is essentially one-sided, so that pulses are emitted in just one direction into the output fiber. We could equally well assume a symmetric cavity, in which case our scheme could be equated to a 50/50 beamsplitter with the incident state in one input port determined by the initial state of the atom. This would provide a straightforward means of producing an entangled state of light fields propagating in opposite directions away from the cavity.

\begin{acknowledgments} 
This work makes use of the Quantum Toolbox in Python (QuTiP) \cite{JOHANSSON20121760,JOHANSSON20131234}. We also acknowledge the contribution of NeSI high-performance computing facilities to the results of this research. New Zealand's national facilities are provided by the New Zealand eScience Infrastructure and funded jointly by NeSI's collaborator institutions and through the Ministry of Business, Innovation and Employment's Research Infrastructure program.
\end{acknowledgments}

\bibliographystyle{apsrev4-1}
\bibliography{bib.bib}% Produces the bibliography via BibTeX.,bib.bib

\onecolumngrid
\newpage
\begin{center}
    \LARGE{\textbf{Supplemental Information}}
\end{center}

\section{Atomic state populations}
In Fig.~\ref{atmpops} below we show the time evolution of the atomic state populations for two of the parameter sets used in Fig.~2 of the main letter that yield pulses of similar duration. For convenience, in the first column we plot the output photon flux again. The atomic population is seen to transfer smoothly and predominantly along the $F=2$ ground state manifold from the initial state $|F=2,m_F=-2\rangle$ to the state $|F=2,m_F=+2\rangle$. The peak in the output photon flux coincides approximately with the peak in the population of the state $|F=2,m_F=-1\rangle$. A small amount of population makes it into the $F=1$ manifold, where it is also subject to effective, resonant (spin-1) anti-Tavis-Cummings dynamics and is transferred to the state $|F=1,m_F=+1\rangle$. From there, a far-off-resonant Raman transition eventually transfers this small population into the target state $|F=2,m_F=+2\rangle$.

For the two cases considered in Fig.~\ref{atmpops} there is not a large difference in the populations of the $F=2$ manifold. The $F=1$ manifold on the other hand shows some slight differences; it depletes more rapidly (even though initially more highly populated) for the nanocavity system as a result of the larger atom-cavity coupling strength, which enhances the off-resonant Raman transition back into the $F=2$ ground state. The figure also underpins our assumptions that the excited state and lower-ground-state populations are very small, thus ensuring the validity of the effective Tavis-Cummings (or anti-Tavis-Cummings) model.

\renewcommand{\thefigure}{S1}
\begin{figure}[h]
\centering
\includegraphics[width=0.3\linewidth]{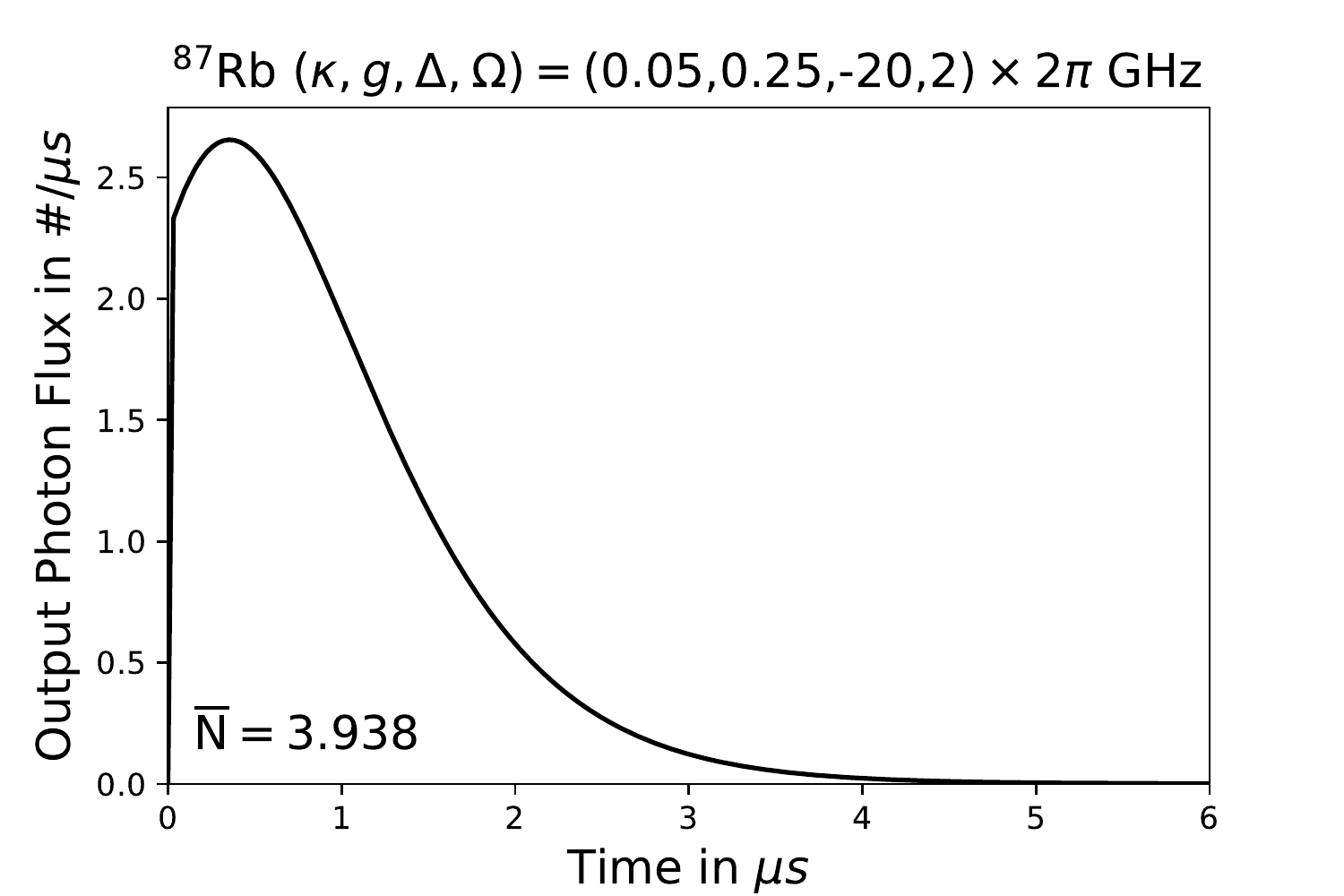}
\includegraphics[width=0.3\linewidth]{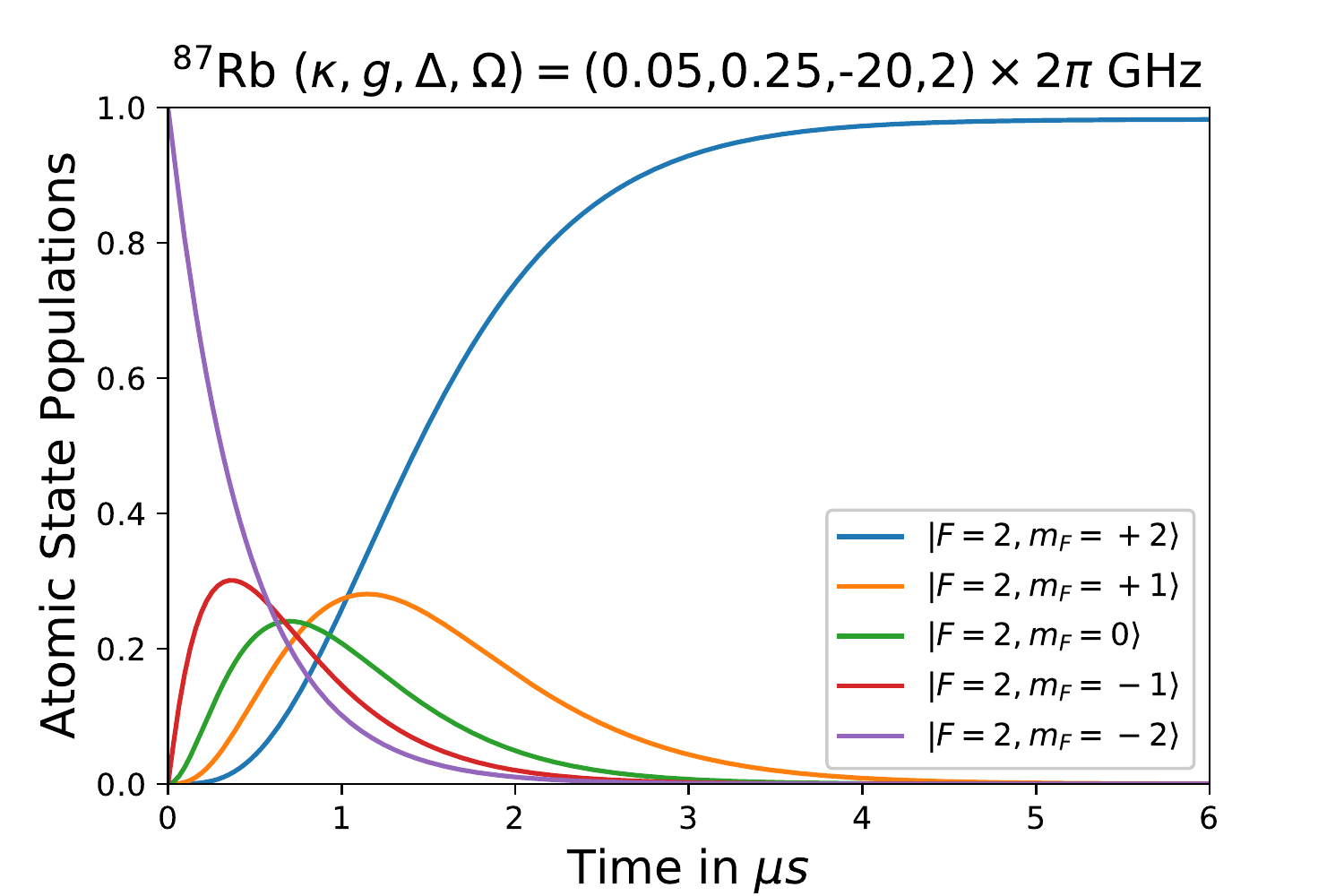}
\includegraphics[width=0.3\linewidth]{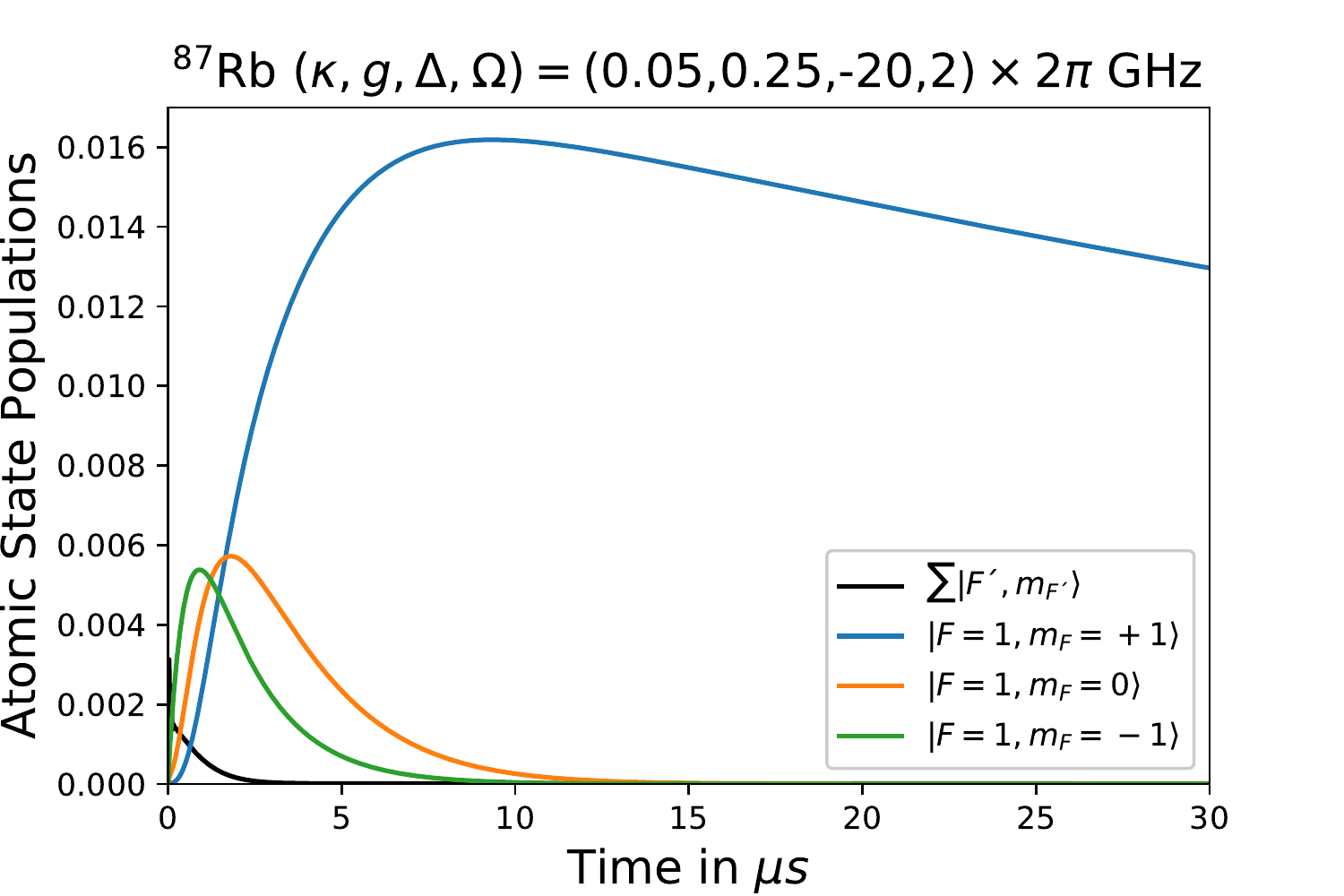}\\
\includegraphics[width=0.3\linewidth]{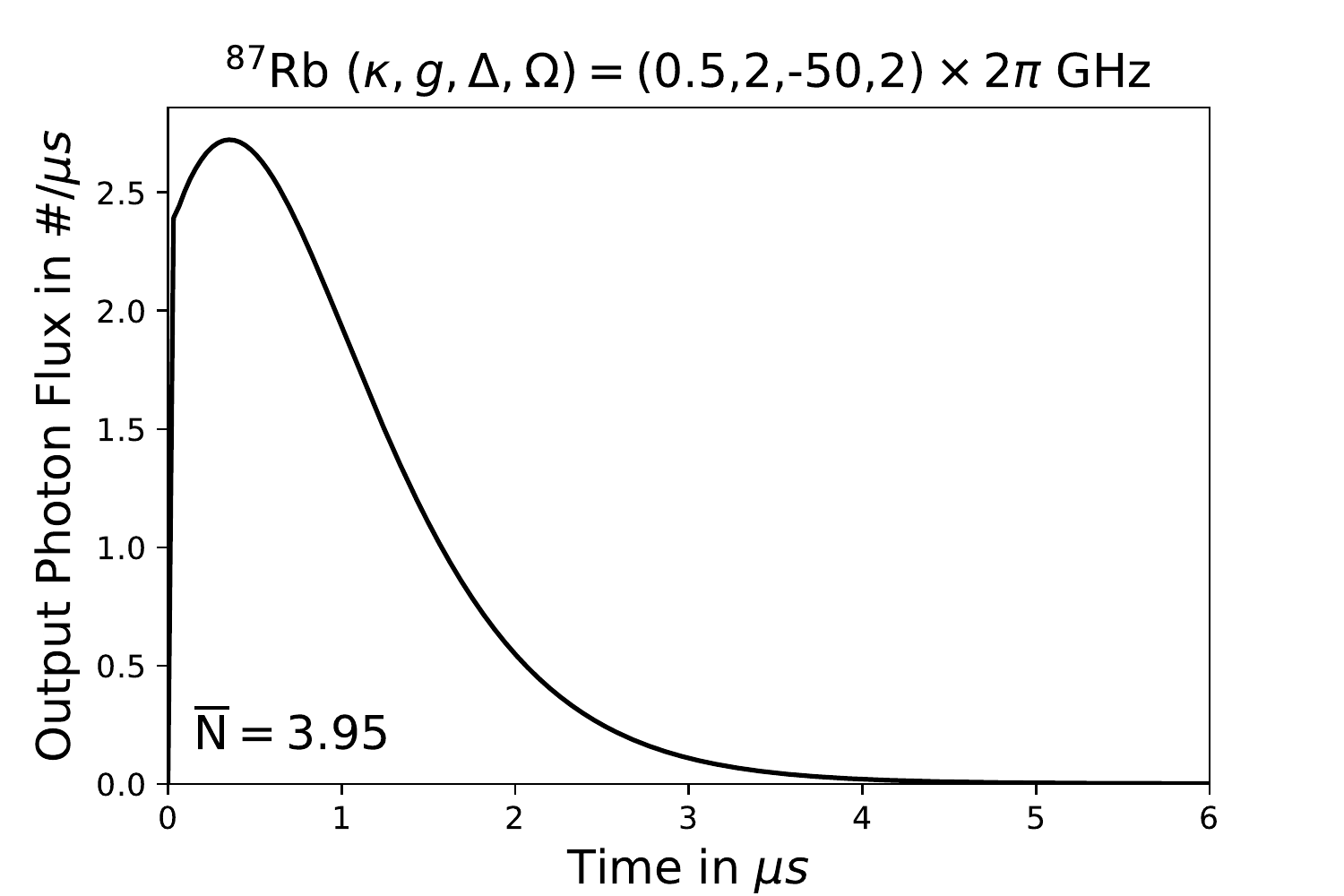}
    \includegraphics[width=0.3\linewidth]{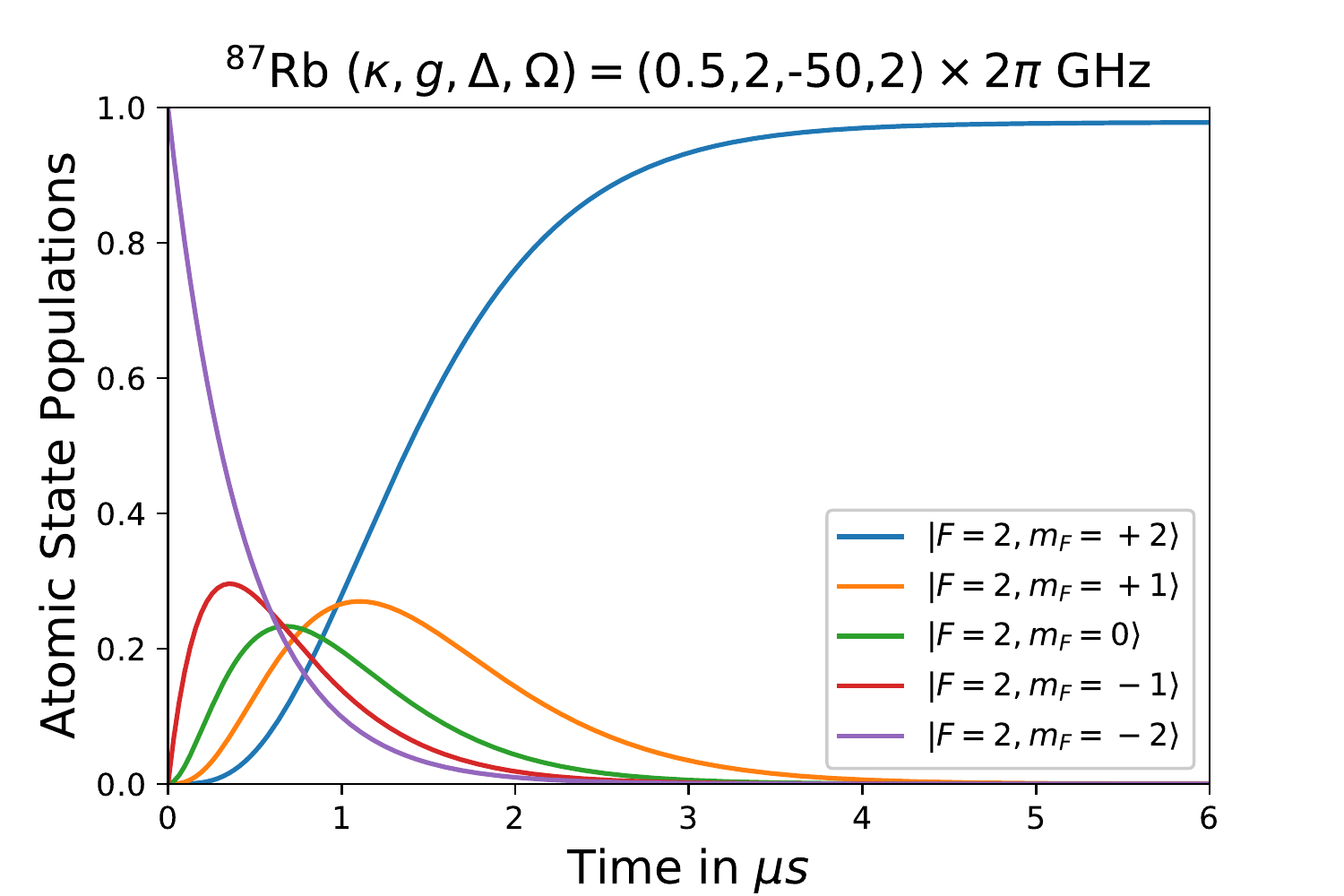}
    \includegraphics[width=0.3\linewidth]{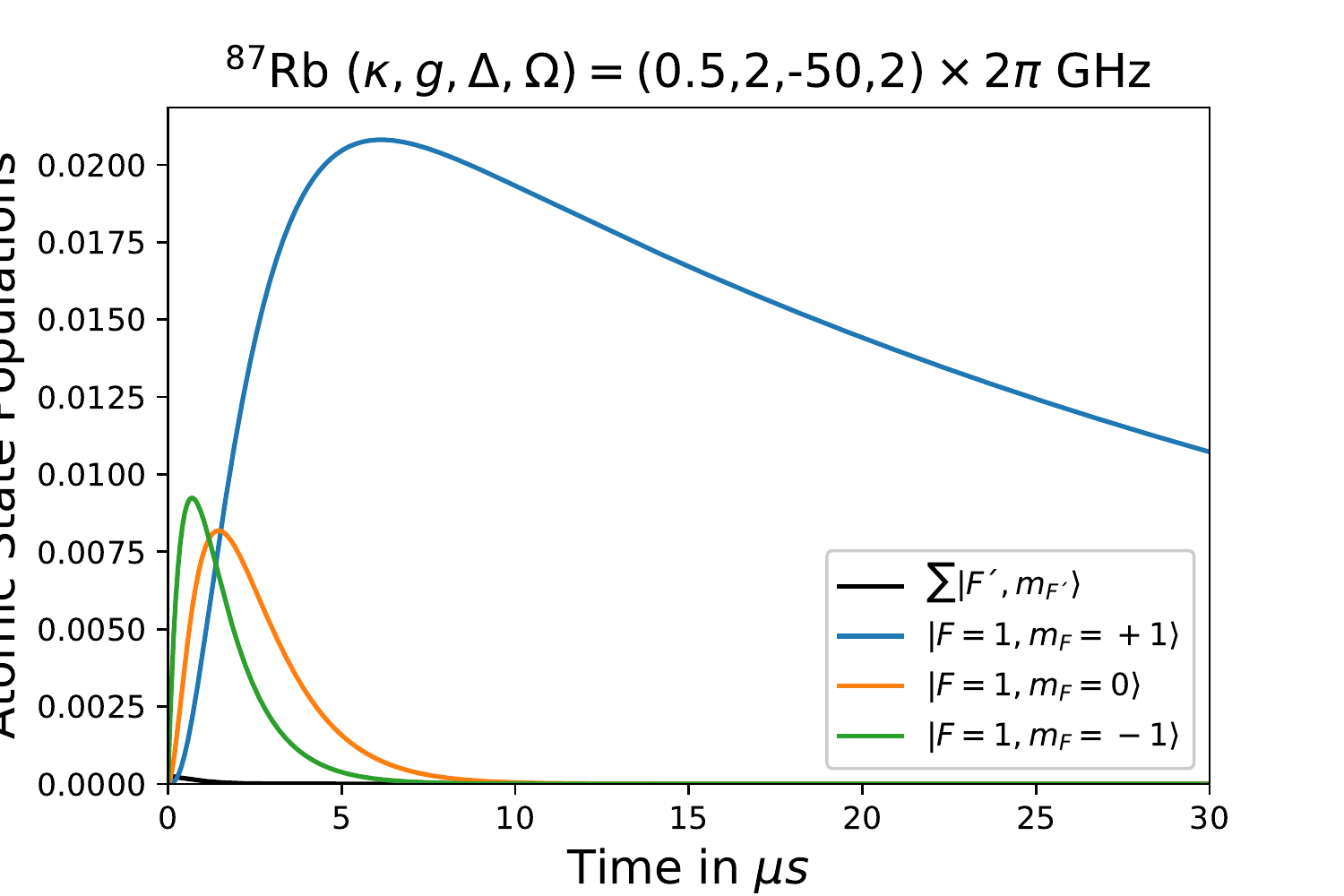}
	\caption{Output photon flux (left column), atomic ground state populations ($F=2$: middle column, $F=1$: right column), and total excited state population (right column) as a function of time for a ${}^{87}{\rm Rb}$ atom initially prepared in state $|F=2,m_F=-2\rangle$. Parameters are for optical microcavity (top row) and nanocavity (bottom row) systems.}
	\label{atmpops}
\end{figure}

\section{Additional Examples}

\subsection{Constant laser amplitude}
As suggested by the formula for the approximate time scale of the output pulse, 
\begin{equation}
\tau\propto\frac{\kappa\Delta^2}{Fg^2|\Omega|^2} ,
\end{equation} 
we can create even shorter pulses by further decreasing the detuning $\Delta$. While the approximation that the detuning is much larger than the excited state hyperfine splitting might still be reasonably valid, at some point we generally observe a discrepancy between the full system dynamics and the Tavis-Cummings dynamics. This is typically in the form of oscillations during the initial rise of the output photon flux, and a small offset in time of the output pulse from that of the simplified model. Some examples are shown in Fig.~\ref{additional_examples}. 

We observe that rapid oscillations in the initial stages of the output pulse are most prevalent at lower detunings for the nanocavity parameter regime, where, in particular, we find that the relative magnitude of $\kappa$ plays a significant role. However, this can be counteracted to a certain degree, or effectively removed, by increasing $\Omega$ and/or $g$ sufficiently.

While the time offset seems to increase by lowering $\Delta$ (see top row of Fig.~2 in the main letter), the change is very small. It is more influenced by the amplitude of $\kappa$ and will persist through to larger values of $\Delta$, as shown in the middle row of Fig.~\ref{additional_examples}, where we note that the cavity decay rate $\kappa$ is a factor 2 larger than in the main text (and actually closer to state-of-the-art values \cite{thompson2013coupling}).

Overall, however, the results shown in Fig.~\ref{additional_examples} demonstrate that the (anti-)Tavis-Cummings model still captures very well the essential behavior of the output pulse even at significantly lower detunings than considered in the main letter, and hence also for significantly shorter pulses.

\renewcommand{\thefigure}{S2}
\begin{figure}[h]
\centering
\includegraphics[width=0.49\linewidth]{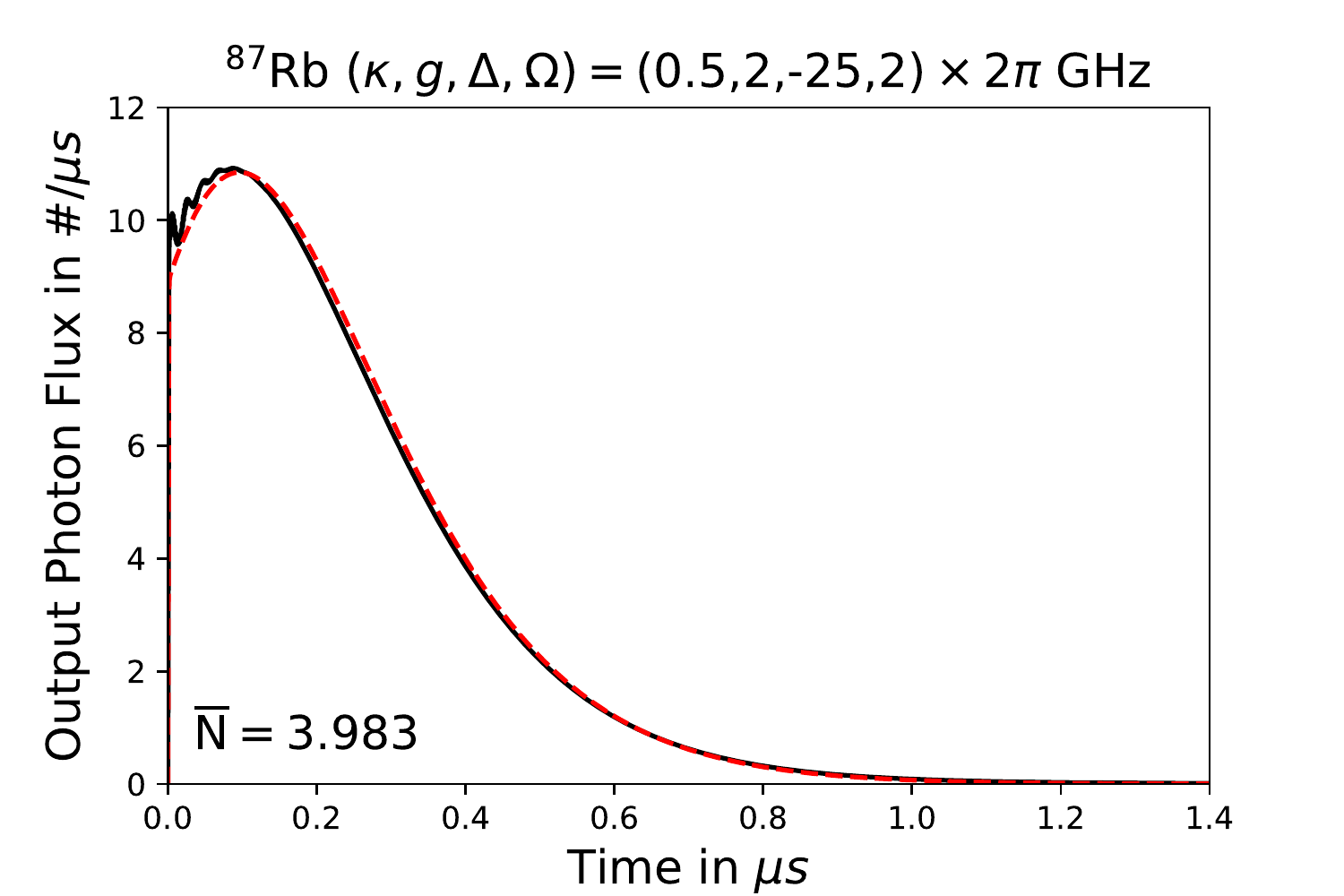}
\includegraphics[width=0.49\linewidth]{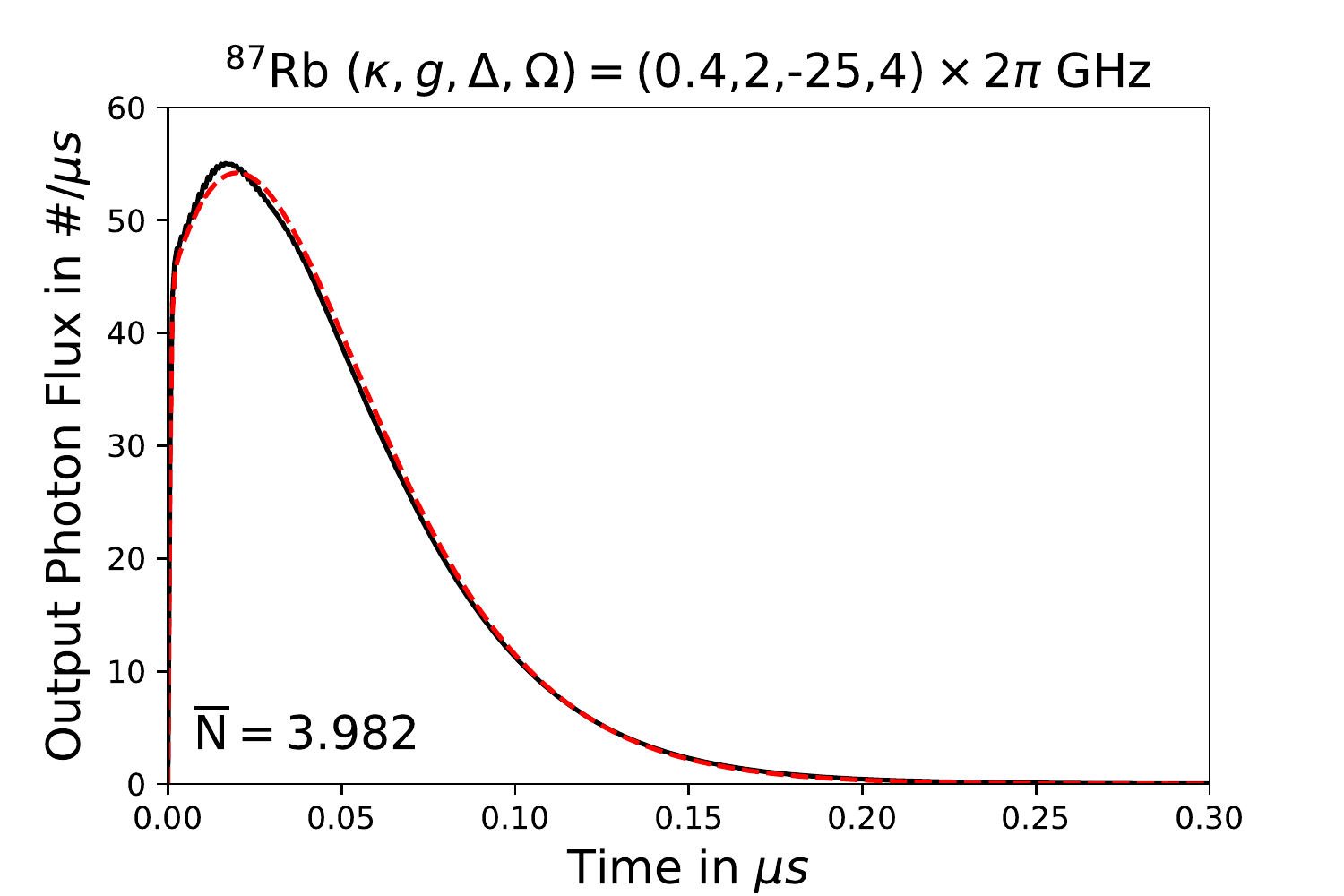}\\
    \includegraphics[width=0.49\linewidth]{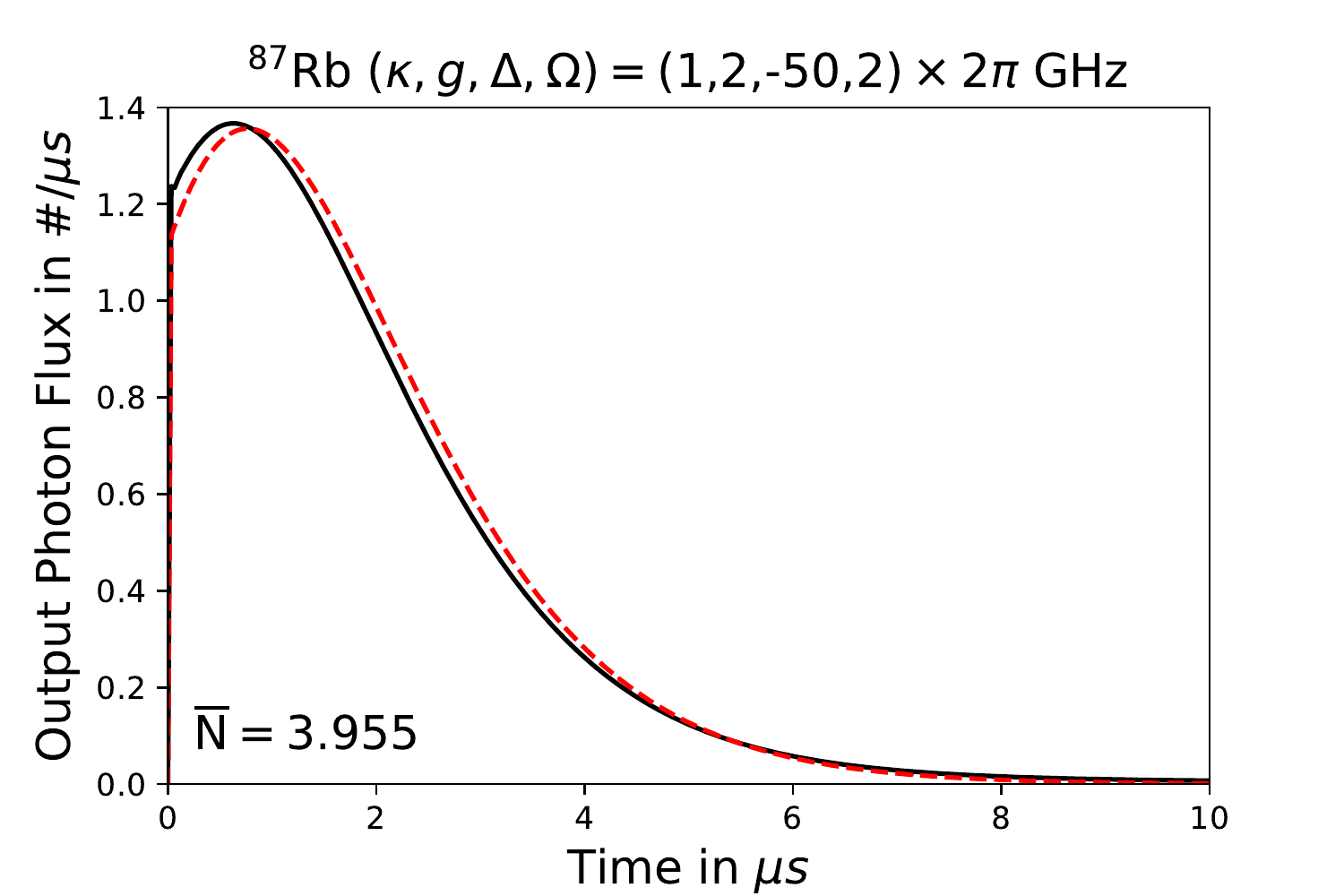}
    \includegraphics[width=0.49\linewidth]{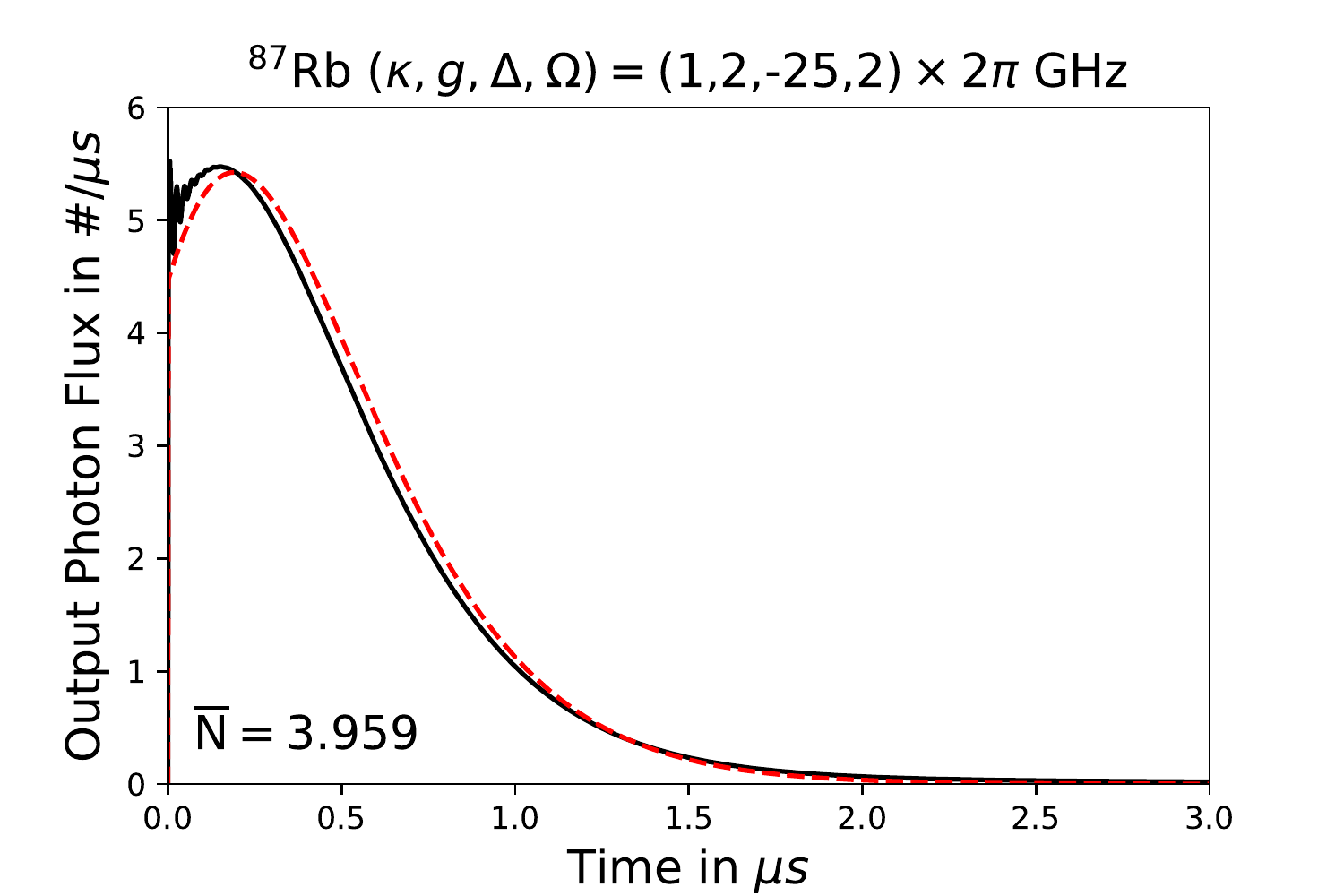}\\
    \includegraphics[width=0.49\linewidth]{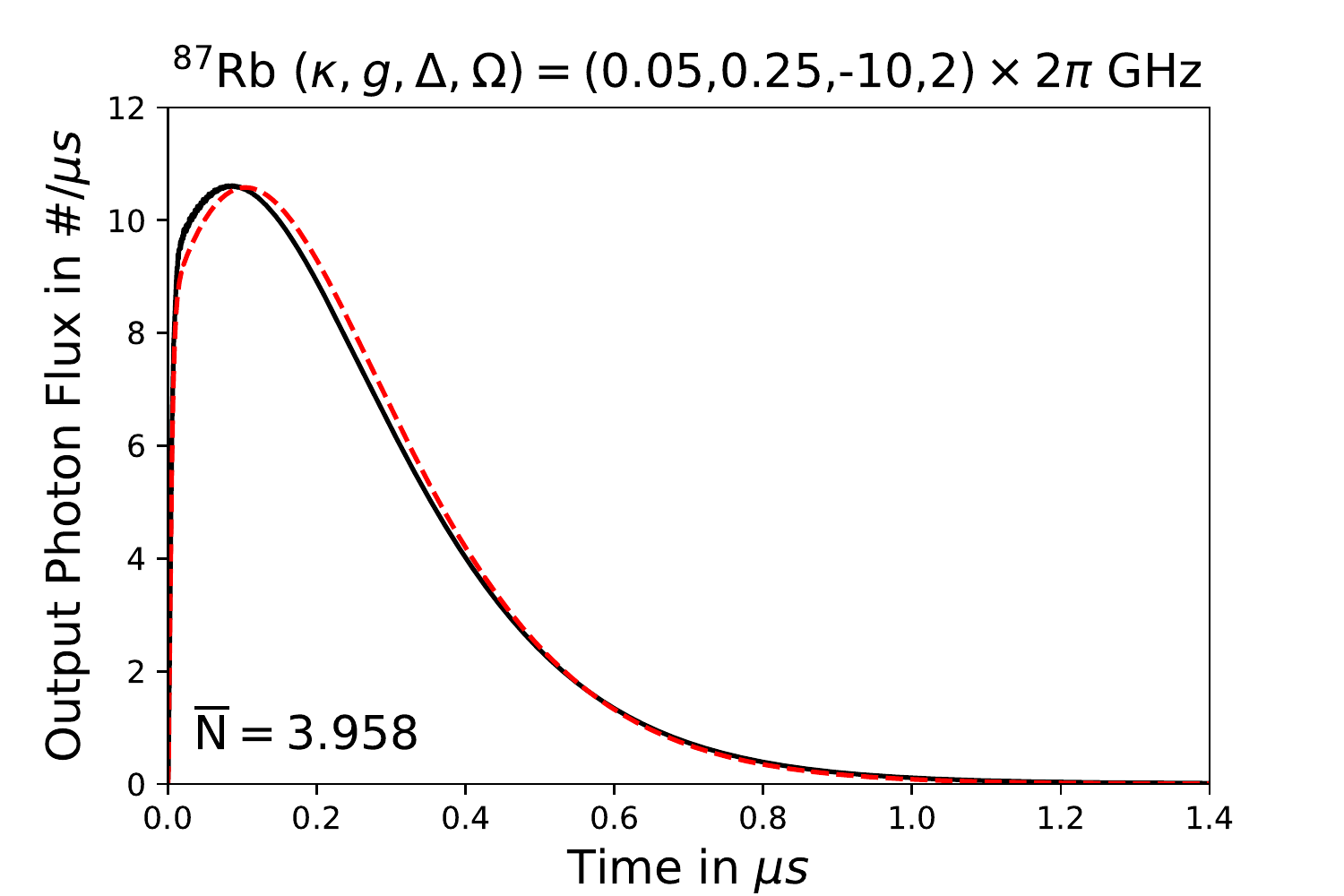}
    \includegraphics[width=0.49\linewidth]{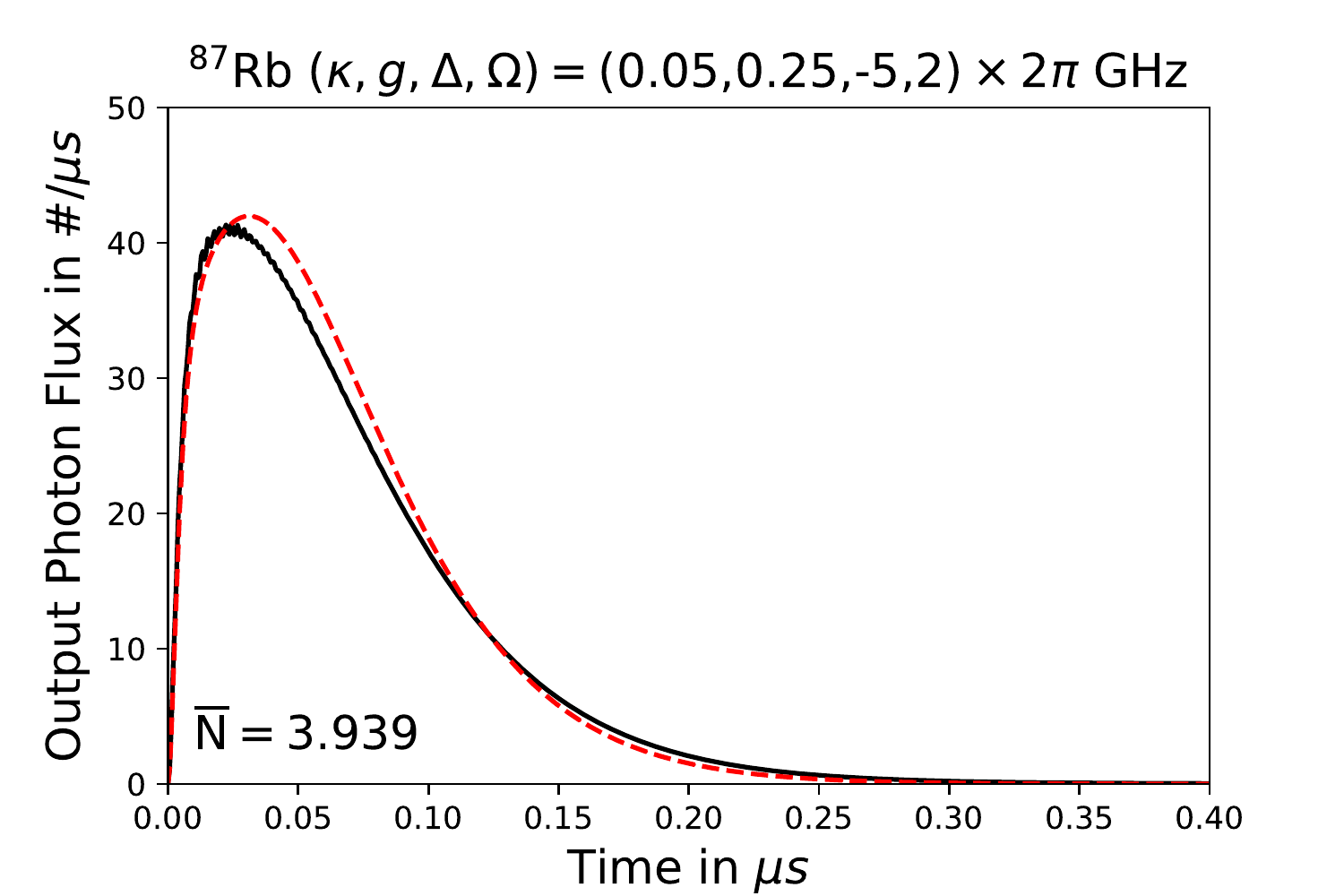}
	\caption{Output photon flux for a ${}^{87}{\rm Rb}$ atom initially prepared in $|F=2,m_F=-2\rangle$. The lines represent the full model (solid black) and the simple Tavis-Cummings model where the excited states have been adiabatically eliminated (dashed red). The number below the curves represents the mean output photon number from the full model.
	}
	\label{additional_examples}
\end{figure}

\newpage
\subsection{Time-varying laser amplitude}

In practice, the laser does not turn on instantaneously at the chosen intensity, but rather increases through a continuous ramp to its final strength. To model this situation, we consider a ramp function of the form
\begin{equation}
r(t)=
\begin{cases}
\sin(\frac{\pi t}{2 t_r}), &t<t_r\\
1, &t>t_r
\end{cases} ,
\end{equation}
and show the resulting output pulses in Fig.~\ref{ramp} for a range of values of the ramp time $t_r$. We note firstly that, for the parameter set considered, ramping the intensity eliminates the rapid oscillations that appear for an instantaneous turn-on of the laser ({\em cf}. Fig.~\ref{additional_examples}). Secondly, for increasing ramp time, the output pulse exhibits an increasingly smooth initial rise, and consequently one is able to produce pulses with both smooth leading and trailing edges. It is clear that by varying the ramp function, one can tailor the temporal profile of the output light pulse. Finally, we note that excellent agreement is maintained between the full model and the simple Tavis-Cummings model with a time-dependent effective coupling strength.

\renewcommand{\thefigure}{S3}
\begin{figure}[t]
\centering
\includegraphics[width=0.49\linewidth]{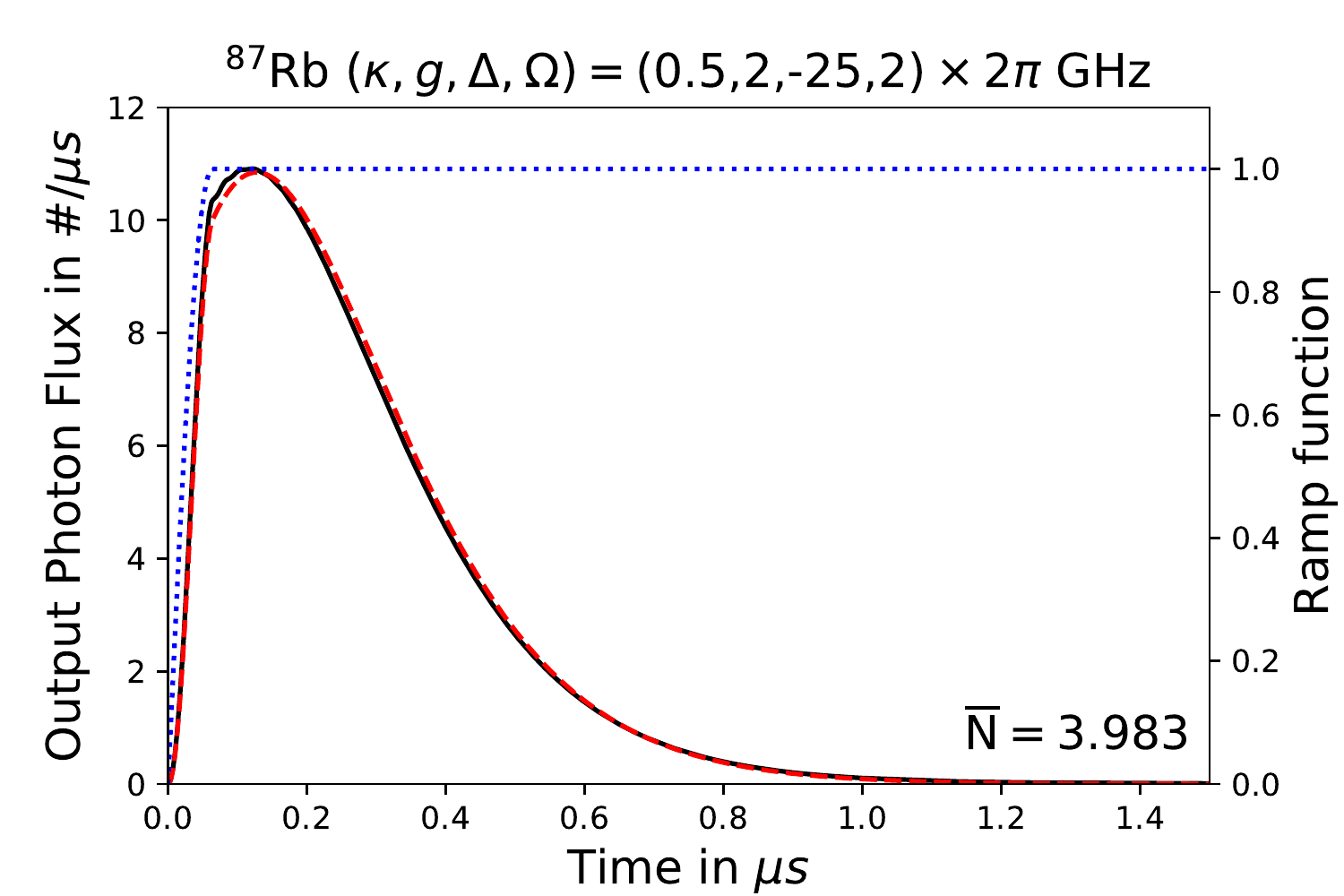}
\includegraphics[width=0.49\linewidth]{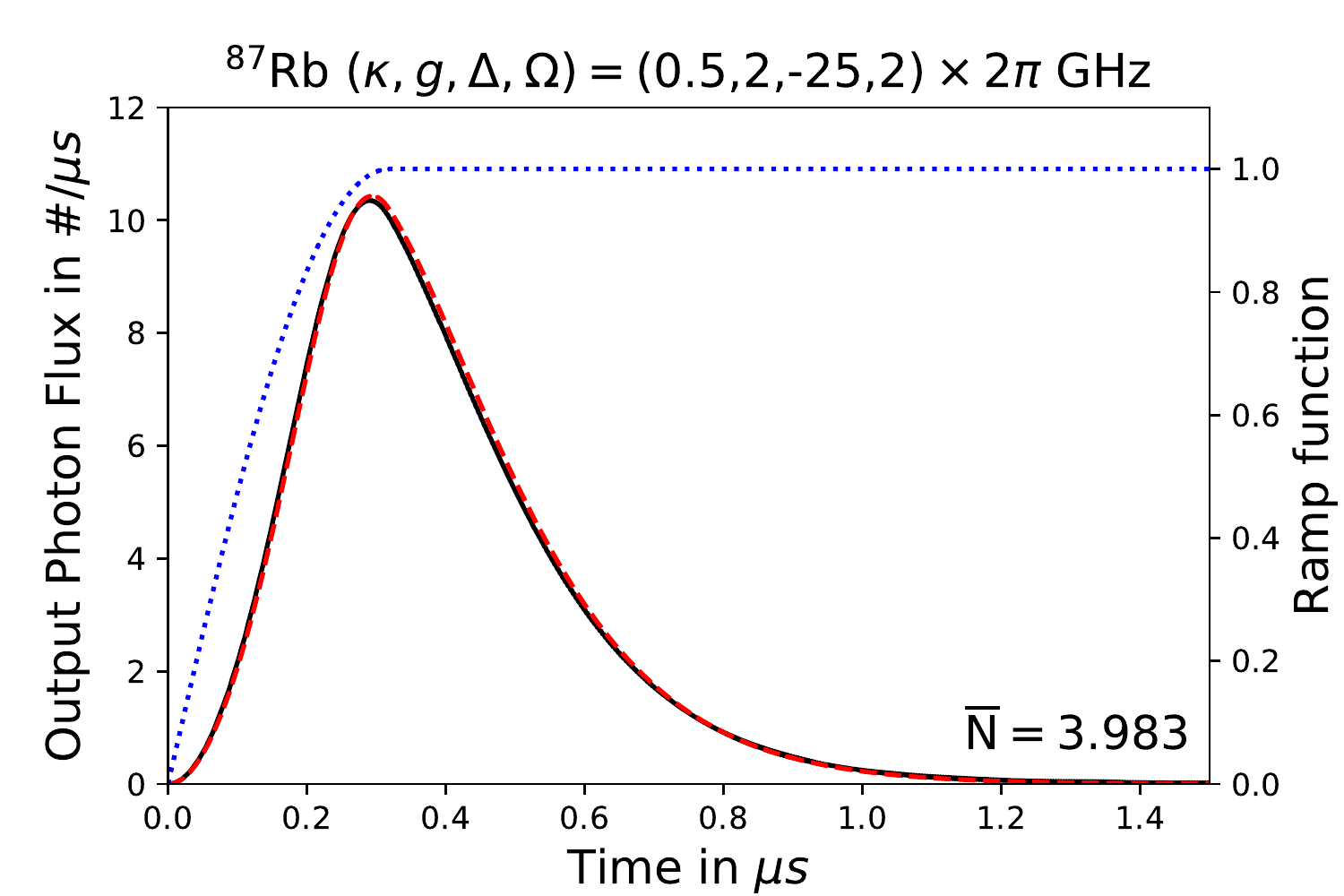}\\
\includegraphics[width=0.49\linewidth]{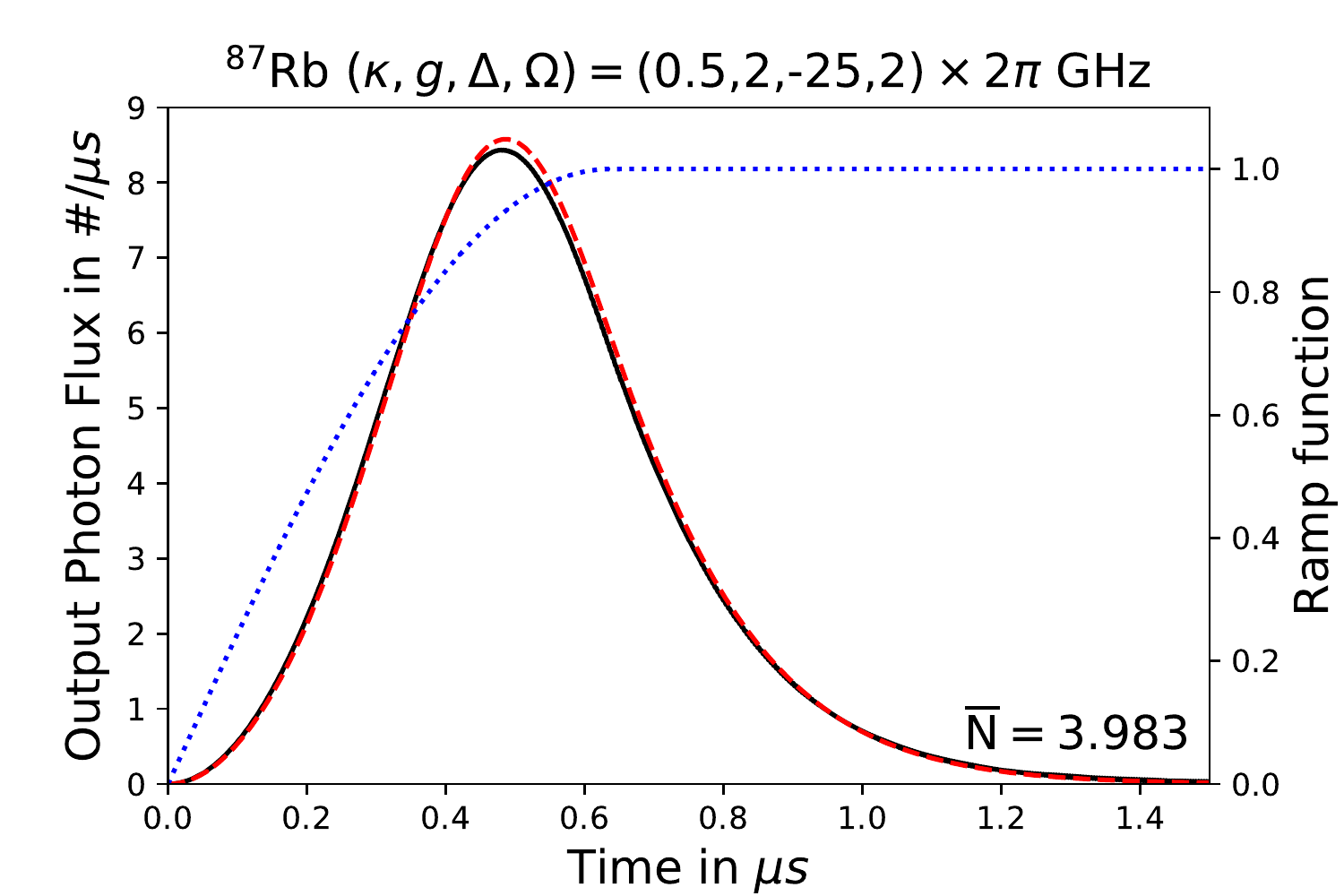}
\includegraphics[width=0.49\linewidth]{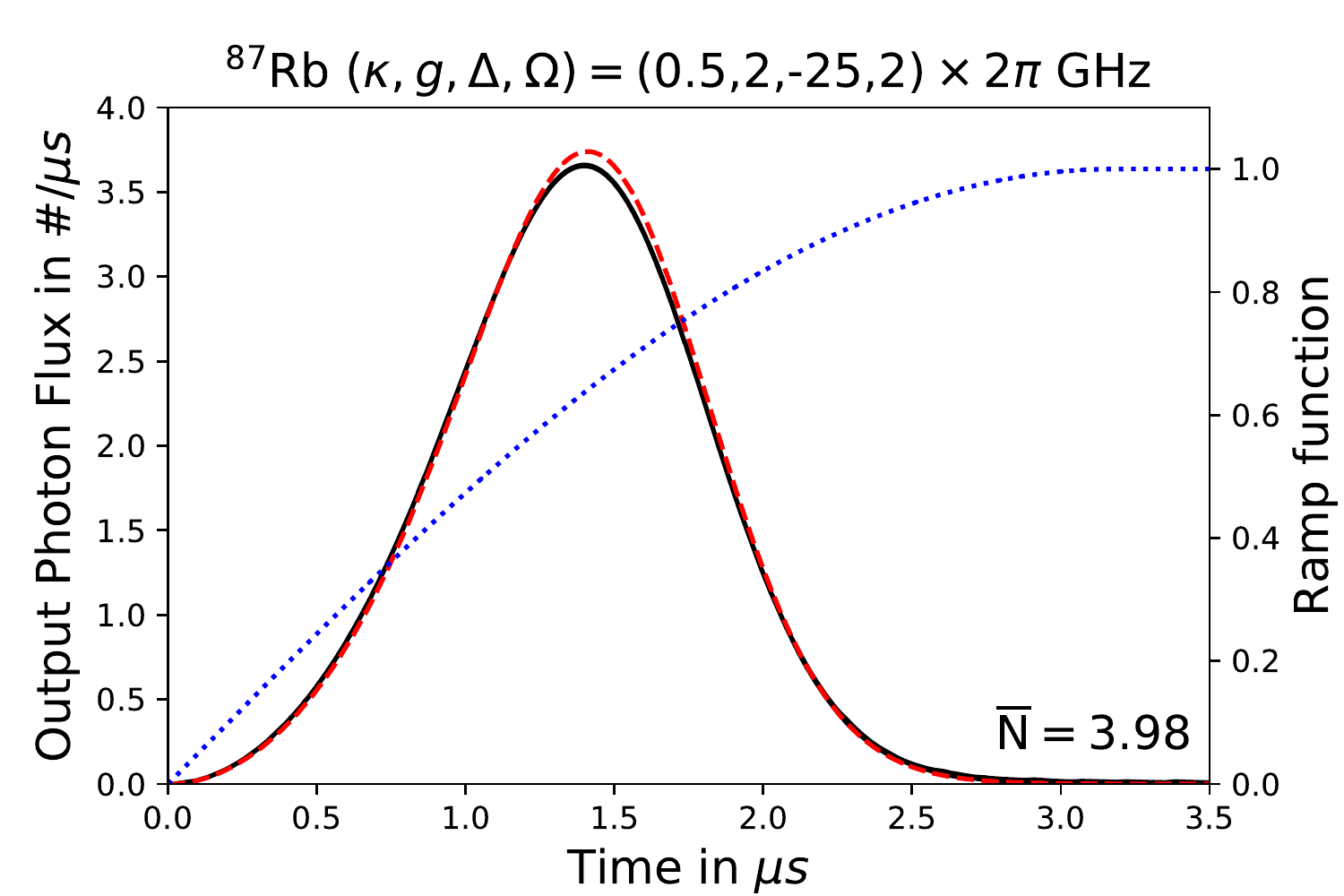}
	\caption{Output photon flux for a ${}^{87}{\rm Rb}$ atom initially prepared in $|F=2,m_F=-2\rangle$. The lines represent the full model (solid black), the simple Tavis-Cummings model where the excited states have been adiabatically eliminated (dashed red) and the ramp function $r(t)$ (dotted blue) that has been applied to both models. The number below the curves represents the mean output photon number from the full model.}
	\label{ramp}
\end{figure}

\section{Quantum Jump Simulations and Modelling Spontaneous Emission in the Effective Tavis-Cummings Model}

\subsection{Quantum Trajectories}
 
To access the variance in the output photon number we perform
photon-counting quantum jump simulations \cite{Molmer:93,carmichael1993open}, where we accumulate photocount records from an ensemble of quantum trajectories. 
In each quantum trajectory, the system state is evolved with the non-Hermitian effective Hamiltonian
\begin{equation}
	\hat H_\text{eff}=\hat {\cal H}_\pm-i\kappa\hat a^\dagger\hat a-i\frac{\gamma}{2} \sum_q  \hat{\mathcal{L}}_q^\dagger\hat{\mathcal{L}}_q ,
\end{equation}
where $\hat{\cal H}_\pm$ is given by Eq.~(5) in the main text and $\hat{\mathcal{L}}_q$ ($q=-1,0,+1$) is an effective spontaneous emission operator (see below).
In particular, over an infinitesimal time step $\Delta t$, we have
\begin{equation}
	|\psi(t+\Delta t)\rangle=e^{-i\hat H_\text{eff}t}|\psi(t)\rangle\approx\left(1-i\hat H_\text{eff}\Delta t+\mathcal{O}(\Delta t^2)\right)|\psi(t)\rangle.
\end{equation}
The loss in norm through this non-Hermitian evolution is given by 
\begin{equation}
	\Delta p=\Delta t\langle\psi(t)|\kappa\hat a^\dagger\hat a|\psi(t)\rangle +
\Delta t\langle\psi(t)|\frac{\gamma}{2}\sum_q\hat{\mathcal{L}}_q^\dagger\hat{\mathcal{L}}_q|\psi(t)\rangle ,
\end{equation}
which corresponds to the probability for a quantum jump to occur. If a jump does occur (as decided by a random number), we replace the effective Hamiltonian evolution with the action of a quantum jump, itself also chosen by a random number: 
\begin{equation}
|\psi(t+\Delta t)\rangle\xrightarrow{}\begin{cases}
&\frac{\sqrt{2\kappa}\hat a|\psi(t)\rangle}{\sqrt{\langle\psi(t)|2\kappa\hat a^\dagger\hat a|\psi(t)\rangle}}~~~ \text{or}\\
&\frac{\sqrt{\gamma}\hat{\mathcal{L}}_q|\psi(t)\rangle}{\sqrt{\langle\psi(t)|\gamma\hat{\mathcal{L}}^\dagger_q\hat{\mathcal{L}}_q|\psi(t)\rangle}} ~~ (q=-1,0~\text{or}~+1).
\end{cases}
\end{equation}
If the jump corresponds to emission of a photon from the cavity, we record its timestamp and increase the photon count for that trajectory by one; sampling over many trajectories then yields the temporal distribution of the output photons and the output photon number histogram, respectively.

\subsection{Spontaneous Emission in the Effective Tavis-Cummings model}

In the quantum jump simulations as just described, the output photon number count can only deviate (i.e., decrease) from the maximal value of $2F$ through atomic spontaneous emission (see Fig.~\ref{SpE}). The adiabatic elimination of the excited states can be extended to incorporate effects of spontaneous emission in a way that is compatible with the effective Tavis-Cummings model. To do this, we use the effective Lindblad operator formalism described in \cite{PhysRevA.85.032111}. Note that the cavity decay remains unchanged. Since the simple Tavis-Cummings model ignores the lower hyperfine ground state, we shall also ignore decay to this state. The new effective spontaneous emission operators,
\begin{equation}
    \hat{\mathcal{L}}_q=\hat L_q\hat{\mathfrak{H}}^{-1}\hat V_+ ,
\end{equation}
include a dissipative part $\hat L_q$ corresponding to the normal spontaneous emission operator for the different polarizations from the full model,
\begin{equation}
    \hat L_q= \sum_{F,F'} \hat D_q(F,F') ,
\end{equation}
and a coherent part $\hat V_+$ corresponding to the interaction part of $\hat H_\pm$ responsible for excitations from the ground state to the excited state manifold, i.e., to the coherent evolution just before the decay:
\begin{equation}
    \hat V_+=\left[\sum_{F,F'}\left( \frac{\Omega}{2} \hat D_{\pm 1}(F,F')  
    +g\hat a^\dagger \hat D_0(F,F') \right)\right]^\dagger.
\end{equation}
Finally, $\hat{\mathfrak{H}}$ is the non-Hermitian Hamiltonian from the quantum trajectory picture \cite{carmichael1993open} that describes the evolution of the excited state
\begin{equation}
    \hat{\mathfrak{H}}=-\sum_{F',m_{F'}} (\Delta_{F'}+\omega_\text{Z}(F',m_{F'}))|F',m_{F'}\rangle\langle F',m_{F'}|-i\frac{\gamma}{2}\sum_q \hat L_q^\dagger \hat L_q\approx-\sum_{F',m_{F'}} \Delta_{F'}|F',m_{F'}\rangle\langle F',m_{F'}|.
\end{equation}

\renewcommand{\thefigure}{S4}
\begin{figure}
\centering
	\includegraphics[width=0.49\linewidth]{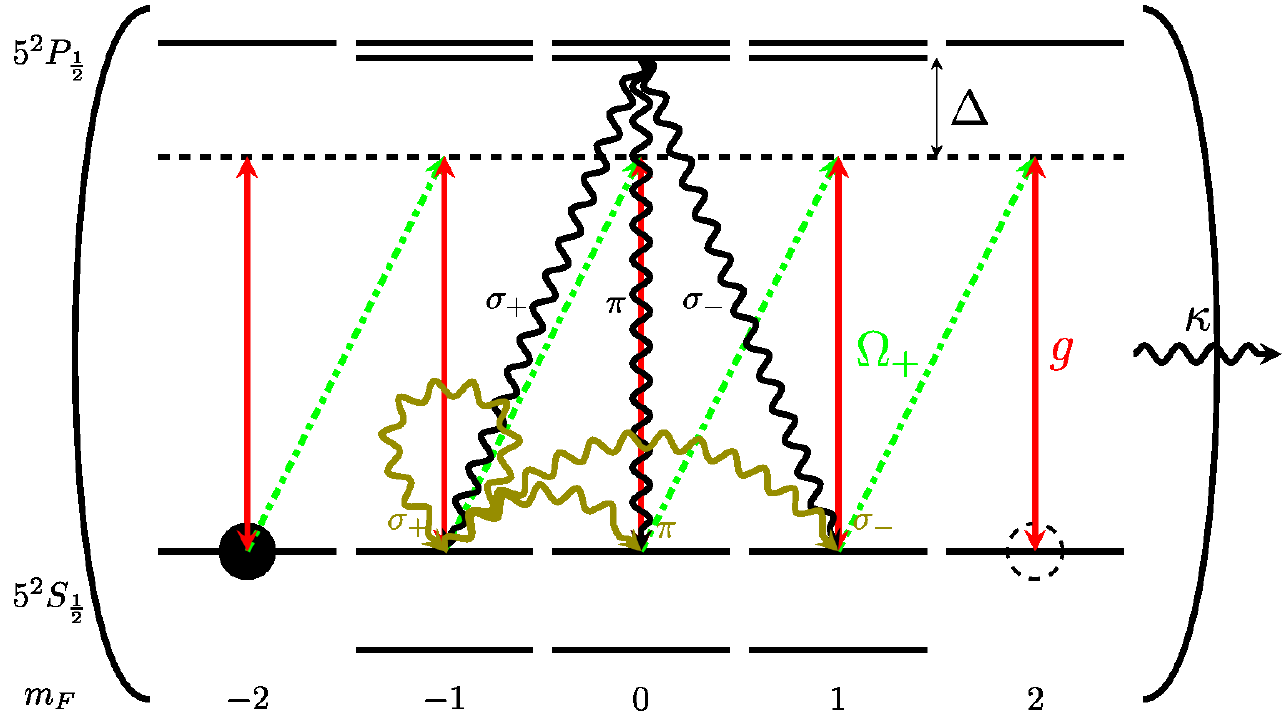}
	\caption{Schematic representation of the action of spontaneous emission in the full model and its effective action in the model reduced to one hyperfine ground state. Depending on the polarization of the emitted photon ($\pi$, $\sigma_-$, or $\sigma_+$), the photon number in the output pulse is reduced by one, two, or zero photons, respectively.}
	\label{SpE}
\end{figure}

\newpage
\section{Balanced Homodyne Detection and Quantum State Reconstruction}

\subsection{Balanced Homodyne Detection}
In balanced homodyne detection \cite{yuen1983noise} the cavity output field is overlapped at a 50/50 beamsplitter with a strong and resonant local oscillator field, i.e., a large coherent state of amplitude $|\epsilon| e^{i\theta}$. The difference in photon counts between the two beamsplitter outputs is then proportional to
\begin{equation}
	\Delta n=|\epsilon|\left(\sqrt{2\kappa}\langle \hat a^\dagger e^{i\theta}+\hat ae^{-i\theta}\rangle\Delta t+\Delta W\right),
\end{equation}
which is directly proportional to the expectation value of the quadrature $\hat x_\theta=\hat a^\dagger e^{i\theta}+\hat a e^{-i\theta}$, while $\Delta W$ is a Wiener noise increment accounting for detector shot noise.

The continuous detection of the output field is modelled with a stochastic Schrödinger equation \cite{carmichael2009statistical} which evolves the wave function between quantum jumps with the propagator
\begin{equation}
	\hat U=1-i\hat H_B\Delta t+\sqrt{2\kappa}\frac{\Delta n}{|\epsilon|}e^{-i\theta}\hat a+\mathcal{O}(\Delta t^2),
\end{equation}
where $\hat H_B$ is the non-Hermitian Hamiltonian from the quantum jump method.
For a detector with bandwidth $\Gamma$, the (scaled) photocurrent obeys
\begin{equation}
	\Delta I(t)=-\Gamma\left( I(t)\Delta t-\frac{\Delta n}{|\epsilon|}\right) .
\end{equation}
For our simulations we consider the bandwidth to be large, e.g., $\Gamma\approx\frac{1}{\Delta t}$.

\subsection{Quantum State Reconstruction}
\subsubsection{Temporal Modes and the Radon Transform}

 We measure marginals of the Wigner function for a set of angles $\theta$ by scanning the phase of the local oscillator. Note that if $\omega=\omega_0\neq0$, we need to introduce an additional time dependence to the local oscillator to match that of the cavity field, so as to maintain the quadrature being measured.
The quantum state of interest corresponding to the output light pulse can be associated with a temporal mode bosonic creation operator \cite{10.1088/1402-4896/ab6153}, 
\begin{equation}
    \hat A^\dagger=\int dt f(t)\hat a^\dagger(t).
\end{equation}
Since we want to mode match to a field amplitude, the natural choice for the filter function is the normalized amplitude correlation function,
\begin{equation}
    f(t)=\frac{\langle \hat a^\dagger(\epsilon)\hat a(t)\rangle}{\sqrt{\int_\epsilon^T dt'\langle \hat a^\dagger(\epsilon)\hat a(t')\rangle^2 }},
    \label{filter}
\end{equation}
which is related to the power spectrum $P(\nu )$ via the Fourier transform
\begin{equation}
    P(\nu)=\int_{-\infty}^{\infty}\langle \hat a^\dagger(\epsilon)\hat a(t)\rangle e^{-i\nu t}dt.
\end{equation}
We start an infinitesimal time $\epsilon$ after $0$, enabling a small increase in cavity excitation, otherwise, because of the initial vacuum state, the filter would just be zero, i.e., $\langle \hat a^\dagger(0)\hat a(t)\rangle=\text{Tr}\{\hat ae^{\mathcal{L}t}[\rho(0)\hat a^\dagger ]\}=0$.
The photocurrent is filtered for the chosen temporal mode
\begin{equation}
	\Delta n=|\epsilon|\left(\sqrt{2\kappa}\langle f(t) \hat a^\dagger e^{i\theta}+ f^*(t)\hat ae^{-i\theta}\rangle\Delta t+\Delta W\right),
\end{equation}
or equivalently, if $f(t)$ is real, we obtain a single measurement of the chosen quadrature by integrating the filtered photocurrent of a single trajectory over time,
\begin{equation}
    \left[x_\theta\right]_i=\int_0^T dt f(t)I(t).
\end{equation}
Experimentally, the filter could be implemented either by tailoring the temporal profile of the local oscillator to match the filter $f(t)$ or by digital post-processing of the measured photocurrent. 

The Radon transform $\mathcal{R}(x_\theta,\theta)$ \cite{PhysRevA.40.2847} of the Wigner distribution $W(x,y)$ is a marginal distribution,
\begin{equation}
\begin{split}
    \mathcal{R}(x_\theta,\theta)[W(x,y)]&=\int_{-\infty}^\infty\int_{-\infty}^\infty W(x,y)\delta(x_\theta-x\cos\theta-y\sin\theta)dx dy\\
    &=\int_{-\infty}^\infty W(x_\theta\cos\theta-y_\theta\sin\theta,x_\theta\sin\theta+y_\theta\cos\theta)dy_\theta\\
    &=\langle x_\theta|\hat \rho|x_\theta\rangle ,
\end{split}  
\end{equation}
and we perform the inverse transformation to reconstruct the Wigner distribution from the marginals.

\subsubsection{Maximum Likelihood Estimation}
With maximum likelihood estimation \cite{PhysRevA.55.R1561,Lvovsky2004} we are trying to find the density operator that is the most likely to generate our set of $N_\theta$ measured marginals. For this, we start with an initial guess $\hat\rho_0=\mathbb{1}_5/5$ (for the case of an effective spin-2 system, where the Hilbert space can be truncated at $|4\rangle$) and iterate like so,
\begin{equation}
    \hat\rho_{k+1}=\frac{\hat R\hat\rho_k\hat R}{\text{Tr}(\hat R\hat\rho_k\hat R)},
\end{equation}
where
\begin{equation}
    \hat R=\frac{1}{N_\theta}\sum_{\theta,x_\theta}\frac{f_\theta(x_\theta)}{p_\theta(x_\theta)}|x_\theta\rangle\langle x_\theta|.
\end{equation}
Here $f_\theta$ is our measured marginal function, the relative frequency with which $x_\theta$ was measured, and $p_\theta(x_\theta)=\text{Tr}(|x_\theta\rangle\langle x_\theta|\hat\rho_k)$ is the probability of our current guess $\hat\rho_k$ to give $x_\theta$. In the Fock-state basis we can write
\begin{equation}
\begin{split}
    |x_\theta\rangle\langle x_\theta|&=\sum \langle n|x_\theta\rangle\langle x_\theta|n'\rangle|n\rangle\langle n'| ,\\
    \langle x_\theta|n\rangle&=\frac{2^{-n/2}}{\sqrt{n!}\pi^\frac{1}{4}}e^{-x_\theta^2/2}e^{-in\theta}H_n(x_\theta),
\end{split}    
\end{equation}
where $H_n(x)$ is the $n$-th Hermite polynomial. The iteration converges towards the ideal density operator $\hat\rho$, because as we approach the ideal state, $\text{Tr}(|x_\theta\rangle\langle x_\theta|\hat\rho)=f_\theta(x_\theta)$, and then because of the closure relation $\hat R=\sum_{x_\theta}|x_\theta\rangle\langle x_\theta|=\mathbb{1}_5$.

\subsubsection{Input-Output Theory for Quantum Pulses}
As an alternative, somewhat more abstract, method to reconstruct the quantum state of the output pulse, we can follow the recent Gedanken-experiment put forward in \cite{PhysRevLett.123.123604}, whereby virtual cavities with time-dependent mirror reflectivities supply input-pulses to, or absorb output-pulses from, a target system. 
In the present context, we can consider the outgoing pulse with profile $f(t)$ from (\ref{filter}) to be incident on a virtual, one-sided cavity with cavity mode annihilation operator $\hat b$ and time-dependent mirror coupling chosen to be
\begin{equation}
    g_f(t)=-\frac{f^*(t)}{\sqrt{\int_0^tdt'|f(t')|^2}} .
\end{equation}
With this choice, 
the state of the pulse is asymptotically mapped onto the state of the virtual cavity mode, and, for example, the Wigner function of this mode is then readily computed. 

This setup is treated as a cascaded quantum system \cite{PhysRevLett.70.2269,PhysRevLett.70.2273}, with the additional intercavity coupling Hamiltonian
\begin{equation}
    \hat H=i\sqrt{\frac{\kappa}{2}}\left( g^*_f(t)\hat a^\dagger\hat b-g_f(t)\hat a\hat b^\dagger\right) ,
\end{equation}
and updated cavity decay,
\begin{equation}
\begin{split}
    \kappa\mathcal{D}[\hat a]\hat\rho \rightarrow &\mathcal{D}[\sqrt{\kappa}\hat a+\frac{g^*_f(t)}{\sqrt{2}}\hat b]\hat\rho\\
    &=\kappa\mathcal{D}[\hat a]\hat\rho+\sqrt{\frac{\kappa}{2}}\left( g_f(t)\mathcal{D}[\hat a,\hat b]+g^*_f(t)\mathcal{D}[\hat b,\hat a]\right) \hat\rho+\frac{|g_f(t)|^2}{2}\mathcal{D}[\hat b]\hat\rho ,
    \end{split}
\end{equation}
where the generalized Lindblad form is
\begin{equation}
   \mathcal{D}[\hat a,\hat b]\hat\rho=2\hat a\hat\rho\hat b^\dagger-\hat\rho\hat a^\dagger\hat b-\hat a^\dagger\hat b\hat\rho.
\end{equation}
Using this approach, we obtain for the Fock state, $0N$-state, and binomial-code state, respectively, the fidelities 0.922, 0.959, and 0.968. 
These are in accord with the fidelities obtained using simulated homodyne measurements and maximum likelihood estimation.

Input-output simulation for the full atomic model can be quite numerically demanding due to the additional ``capture'' cavity mode. We check that the desired state leaves the cavity also in the full model by looking at the simplest case, a 1-photon pulse from a ${}^{87}{\rm Rb}$ atom initially prepared in $|F=2,m_F=+1\rangle$. We manage to retrieve the 1-photon Fock state in the virtual cavity with a fidelity of 0.997.
As can be seen in Fig.~\ref{filterplot}, in that case, the filter function of the effective model corresponds to the absolute value of the full-model filter function. The non-zero free Hamiltonian of the full model (involving the large detuning $\Delta$) adds a time-dependent phase. 

\renewcommand{\thefigure}{S5}
\begin{figure}[H]
\centering
	\includegraphics[width=0.49\linewidth]{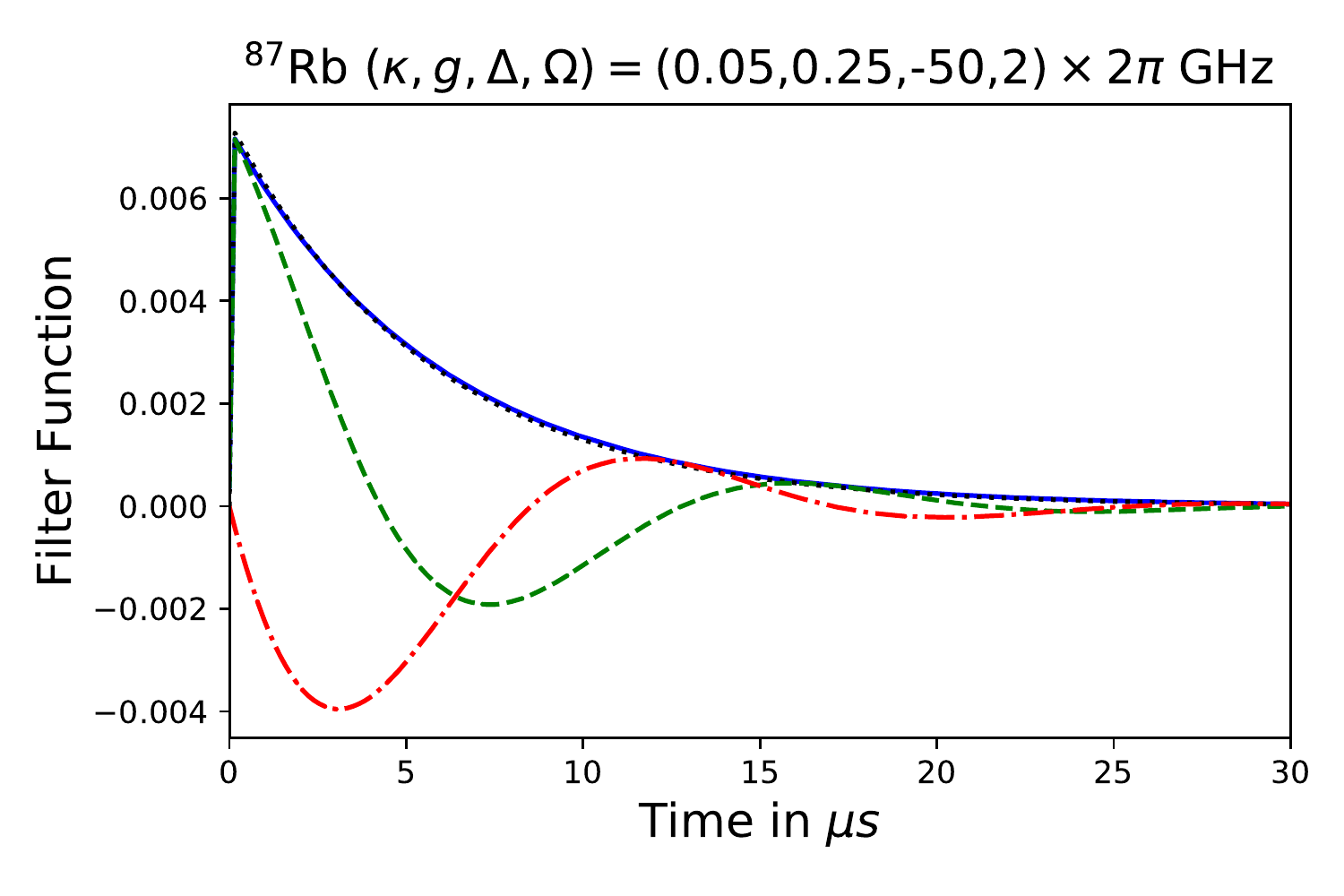}
	\caption{Plot comparing the filter functions for the full and the effective model for a ${}^{87}{\rm Rb}$ atom initially prepared in the ground state $|F=2,m_F=+1\rangle$. The lines represent $|f(t)|$ (solid blue), $\text{Re}(f(t))$ (green dashed), and $\text{Im}(f(t))$ (red dash-dotted) from the full model, and $f(t)$ (black dotted) from the simple Tavis-Cummings model where the excited states have been adiabatically eliminated.}
	\label{filterplot}
\end{figure}

\end{document}